\documentclass[epj]{svjour}
\usepackage{eqnarray,amsmath}
\usepackage{float}
\usepackage{xcolor}
\usepackage{mathtools}
\usepackage{cancel}
\usepackage{appendix}
\usepackage{graphicx}
\usepackage{braket}
\makeatletter
\def\@seccntformat#1{\@ifundefined{#1@cntformat}%
   {\csname the#1\endcsname\quad}
   {\csname #1@cntformat\endcsname}
} \makeatother
\usepackage{amssymb}
\begin{document}
\title{Halo EFT calculation of charge form factor for
two-neutron ${}^{6}\textrm{He}$ halo nucleus: two-body resonant
P-wave interaction}

\author{S. Jesri\inst{1}
\thanks{\emph{e-mail:} sounya.jesri@ut.ac.ir} 
M. Moeini Arani\inst{1}
\thanks{\emph{e-mail:} m.moeini.a@ut.ac.ir} 
S. Bayegan\inst{1}
\thanks{\emph{e-mail:} bayegan@ut.ac.ir(corresponding author)}}%

\institute{Department of Physics, University of Tehran, P.O.Box
14395-547, Tehran, Iran}

\date{Received: date / Revised version: date}
%
\abstract{We take a new look at $^{6}\textrm{He}$ halo nucleus and
set up a halo effective field theory at low energies to calculate
the charge form factor of ${}^{6}\textrm{He}$ system with resonant
P-wave interaction. P-wave Lagrangian has been introduced and the
charge form factor of ${}^{6}\textrm{He}$ halo nucleus has been
obtained at Leading-Order. In this study, the mean-square charge radius of
${}^{6}\textrm{He}$ nucleus relative to ${}^{4}\textrm{He}$ core and
the root-mean-square (r.m.s) charge radius of ${}^{6}\textrm{He}$ nucleus have been
estimated as $\langle r_{E}^{2}\rangle=1.408\; \textrm{fm}^{2}$ and
$\langle r_{E}^{2}\rangle_{{}^{6}He}^{\frac{1}{2}}=2.058
\;\textrm{fm}$, respectively. We have compared our results with the
other available theoretical and experimental data.
 \PACS{
      {11.10.-z}{Field theory}   \and
      {13.40.Gp}{Electromagnetic form factors}   \and
      {21.45.+v}{Few-body systems}
     } 
\keywords{Halo Effective Field Theory, ${}^{6}\textrm{He}$ Halo
nucleus, Charge form factor,Charge radius}
} 
\titlerunning{Halo EFT calculation of charge form factor for
two-neutron ${}^{6}\textrm{He}$ halo nucleus}
\authorrunning{S. Jesri, et al.}
\maketitle
\section{Introduction}\label{introduction}

 Weinberg was the first one who applied the effective field theory
(EFT) to nuclear forces \cite{h12}. Also the concept of applying EFT to nuclear forces was brought by Rho \cite{h41} and by Ord\'{o}\~{n}ez and van Kolck \cite{h42}. An effective field theory
includes the appropriate degrees of freedom to describe physical
phenomena occurring at a chosen length scale or energy scale. Up to
now, cold atoms and few-nucleon systems at low energies have been
studied by this formalism \cite{h13,h14,h15}. Pion-exchange effects
are not resolved at low energies and momenta
($E<\frac{m_{\pi}^{2}}{M_{N}},\; p\leq m_{\pi}$, where $m_{\pi}$ and
${M_{N}}$ are the mass of pion and nucleon respectively), so this
theory is constructed by only short-range contact interactions known
as pionless effective field theory
$(\textrm{EFT}(\pi\mkern-9.5mu/))$ \cite{h16,h17,h18}.

One of the major challenges for nuclear theory is the calculation of
properties of halo nuclei. These nuclei are characterized by a
tightly bound core and one or two weakly bound valence nucleons
\cite{h19,h20,h21,h22}. In halo EFT, either core or nucleons are
treated as the fundamental fields and one can find relations between
different nuclear low-energy observables in this EFT. On the other
hand, most systems can be explained by a short-range EFT expanding
in $\frac{R}{a}$, which $R$ is the range of the nucleon-nucleus
interaction such that $M_{high}\sim\frac{1}{R}$ and $a$ is the
two-body scattering length such that $M_{low}\sim\frac{1}{a}$, so it
is found that $R\ll a$. Based on an EFT in terms of the expansion
parameter $\frac{R}{|a|}$, 2n halo nuclei are described as an
effective three-body system including of a core and two
weakly-attached valence neutrons. Some universal properties of these
nuclei are investigated such as the matter density form factors and
mean square radii \cite{h23,h24}.

Investigations have been carried out into Borromean nuclei such as
${}^{11}\textrm{Li}$, ${}^{14}\textrm{Be}$  ,and ${}^{6}\textrm{He}$
\cite{h19,h20,h23,h24}. While these nuclei have only one bound
state, there are no bound states in the binary subsystems.
Properties of one neutron halo nucleus ${}^{8}\textrm{Li}$ has been
pursued  by the two-body sector halo EFT \cite{h25,h26,h46}. A detailed
analysis of electromagnetic properties of halo nuclei
${}^{11}\textrm{Be}$ system investigated where the low-energy E1
strength function in breakup to the ${}^{10}\textrm{Be}$-neutron
channel has been performed \cite{h36}.

The lightest nuclei with 2n-halo structure are ${}^{6}\textrm{He}$,
${}^{11}\textrm{Li}$, ${}^{14}\textrm{Be}$, ${}^{22}\textrm{C}$, and
${}^{17}\textrm{B}$. Hagen et al. \cite{h8} and Vanasse \cite{h28}
have studied the S-wave EFT framework for ${}^{11}\textrm{Li}$,
${}^{14}\textrm{Be}$, and ${}^{22}\textrm{C}$ nuclei and calculated
corresponding charge radius. Other two-neutron halo nucleus,
${}^{17}\textrm{B}$, will be dealt with in the future works. The
halo EFT we construct includes the two-body P-wave interaction in
the $n\alpha$ subsystem of ${}^{6}\textrm{He}$ nucleus. Binding
energies, radii and other properties of various halo nuclei of
s-wave and p-wave type have been reviewed in halo EFT \cite{h44}. Ji
$\it et al.$ have considered a halo EFT for the three-body
$nn\alpha$ system to explain the ${}^{6}\textrm{He}$ ground state
\cite{h4}. An EFT with P-wave resonance interactions has been
developed for elastic $n\alpha$ scattering by Bertulani et al.
\cite{h1}. Bedaque $\it et al.$ \cite{h43} have suggested a
different power counting compared to \cite{h1} to describe narrow
resonances in EFT and illustrated their results in the case of
nucleon-alpha scattering. The electric dipole strength function
distribution of the ${}^{6}\textrm{He}$ halo nucleus has been
recently evaluated based on the halo EFT approach using the
particle-dimer scattering amplitudes in the $nn\alpha$ system and
the normalized ${}^{6}\textrm{He}$ wave function \cite{h3}. Finally,
the momentum-space probability density of ${}^{6}\textrm{He}$ at
leading order in halo EFT has been presented. The momentum
distribution of ${}^{6}\textrm{He}$ requires the n-n and n-core
t-matrices as well as a c-n-n force as input in the Faddeev
equations \cite{h37}.

In this paper, we focus on the two-neutron halo nucleus
${}^{6}\textrm{He}$, calculate the electric charge form factor and
find the root-mean-square (r.m.s) charge radius of ${}^{6}\textrm{He}$ nucleus. Therefore we
introduce the strong Lagrangian including $n\alpha$ P-wave
interaction for the halo EFT at leading order in Section~\ref{Strong
interaction}. In Section~\ref{Two-body and three-body systems}, the
formalism for two- and three-body propagators are completely
presented. ${}^{6}\textrm{He}$ charge form factor is evaluated in
Section~\ref{Charge form factor}. In Section~\ref{Numerical
calculation and results}, our numerical results for the form factor
and the charge radius are presented and compared with experimental
data. Finally, we conclude in Section~\ref{Conclusion}. In
Appendix~\ref{Appendix A}, the particle-dimer scattering is
explained and in Appendix~\ref{Appendix B}, some expressions for the
contributions of diagrams participate into the form factor of
${}^{6}\textrm{He}$ are presented in details.

\section{Strong interaction}\label{Strong interaction}
\subsection{Power-counting}\label{Power-counting}
We apply a halo effective field theory in non-relativistic formalism
for the alpha core ($\alpha$) with spin zero interacting with two
spin half neutrons. In this method, we define $Q$ as a low momentum
scale attributed to core and neutron momentums. Furthermore, the
high momentum parameter, $\bar{\Lambda}$ can be scaled as
$\bar{\Lambda}\sim m_{\pi}\sim\sqrt{m_{\alpha}E_{\alpha}}$ where
$m_{\alpha}$ and $E_{\alpha}=20.21\;\textrm{MeV}$ refer to the mass
and the excitation energy of $\alpha$ particle.
 There are the two-body neutron-neutron ($nn$) and the neutron-alpha
($n\alpha$) interactions in ${}^{6}\textrm{He}$ calculation. The
remarkable state in the $nn$ is $S$-wave virtual bound state. A
low-momentum scale $Q$ is defined by the inverse of the di-neutron
scattering length, $\frac{1}{(a_{0}=-23.714)}\;\textrm{fm}^{-1}$ and
the inverse of the effective range of this S-wave state,
$\frac{1}{(r_{0}=2.73)}\;\textrm{fm}^{-1}$ is considered as the
high-momentum scale $\bar{\Lambda}$. So, the leading order (LO)
scattering amplitude of two neutrons is constructed by the
scattering length contribution only. With respect to these scales,
the $S$-wave effective range expansion (ERE) for di-neutron system
at the lowest-order can be given by
\begin{eqnarray}
 k\cot\delta_{0}=-\frac{1}{a_{0}}+\cdots.
\label{e1}
\end{eqnarray}

In low-energy region, only $S$- and $P$-wave interactions are
significant in the $n\alpha$ system. There are three possible
partial waves for the $n\alpha$ system, ${}^{2}S_{1/2}$,
${}^{2}P_{1/2}$ and ${}^{2}P_{3/2}$. We use the power counting
introduced by Bertulani et al. in Ref. \cite{h1} which also applied
in the Gamow shell model calculation of ${}^{6}\textrm{He}$ in halo
EFT \cite{h5}. This power counting specifies that $n\alpha$
interaction gets the LO contributions only from both scattering
length and effective range of ${}^{2}P_{3/2}$ channel as
\begin{eqnarray}
\frac{1}{a_{1}}\sim Q^{3}\;\;,\;\;\frac{r_{1}}{2}\sim
Q\;\;,\;\;\frac{\mathcal{P}_{1}}{4}\sim\frac{1}{\bar{\Lambda}},
 \label{e2}
\end{eqnarray}
where $a_{1}=-62.95 \;\textrm{fm}^{3}$, $r_{1}=-0.88\;
\textrm{fm}^{-1}$ and $ \mathcal{P}_{1}=-3.0 \;\textrm{fm}$ are the
scattering length, the effective range and the shape parameter of
${}^{2}P_{3/2}$ state, respectively \cite{h2}. Therefore the
lowest-order terms of effective range expansion for the resonant
P-wave $n\alpha$ system are given by
\begin{eqnarray}
 k^{3}\cot\delta_{1}=-\frac{1}{a_{1}}+\frac{r_{1}}{2}k^{2}+\cdots.
\label{e3}
\end{eqnarray}

\subsection{Lagrangians}\label{Lagrangians}
Generally, the effective field theory expansion parameter is defined
by the momentum ratio $Q/\bar{\Lambda}$ and it creates the
order-by-order pattern of convergence. At LO, the Lagrangian for
${}^{6}\textrm{He}$ system can be written as the summation of one-,
two- and three-body contributions,
$\mathcal{L}=\mathcal{L}^{(1)}+\mathcal{L}^{(2)}+\mathcal{L}^{(3)}$
, where
\begin{eqnarray}
 \mathcal{L}^{(1)}&=&n^{\dag}\Big(i\partial_{0}
 +\frac{\vec{\nabla}^{2}}{2m_{n}}\Big)n+\phi^{\dag}\Big(i\partial_{0}+\frac{\vec{\nabla}^{2}}{2m_{\alpha}}\Big)\phi\nonumber\\
 \mathcal{L}^{(2)}&=&\Delta_{0}d_{0}^{\dag}d_{0}-
\frac{g_{0}}{\sqrt{8}}
\Big(d_{0}^{\dag}(n^{\dag}i\sigma_{2}n)+h.c.\Big)
\nonumber\\&+&\eta_{1}d^{\dag}_{1}\Big[i\partial_{0}+\frac{\vec{\nabla}^{2}}{2(m_{n}+m_{\alpha})}-\Delta_{1}\Big]d_{1}
\nonumber\\
&+&\frac{g_{1}}{2}\Big[d^{\dag}_{1}\vec{S}^{\dag}\cdot[n\vec{\nabla}\phi-(\vec{\nabla}n)\phi]+h.c.
\nonumber\\&-&r
[d^{\dag}_{1}\vec{S}^{\dag}\cdot{\vec{\nabla}(n\phi)}+h.c.]\Big]\nonumber\\
\mathcal{L}^{(3)}&=&\Omega
t^{\dag}t-h\frac{\sqrt{3m_{n}}g_{1}}{2\Lambda}
\Big[t^{\dag}(d_{1}(i\sigma_{2})\vec{S}\cdot\vec{P}n)+h.c.\Big],
\label{e4}
\end{eqnarray}
where $m_{n}$ is the neutron mass and $n,\phi,d_{0}\;(d_{1})$ denote
the two component spinor field of the neutron, the bosonic alpha
core field, the auxiliary dimer field of $nn$ ($n\alpha$) system.
Also, $t$ implies a spin-0 trimer auxiliary field. Moreover we have
\begin{eqnarray}
 \vec{P}=\frac{\tilde{\mu}}{\overrightarrow{m}}i\overleftarrow{\nabla}-\frac{\tilde{\mu}}{\overleftarrow{m}}i\overrightarrow{\nabla},
\end{eqnarray}
that $\overrightarrow{m}$ ($\overleftarrow{m}$)implies the mass of
$n\;(d_{1})$ field and $\tilde{\mu}$ denotes the reduced mass of
$n-d_{1}$ system. $\eta_{1}$ is equal to $\pm1$ ,
$g_{0}^{2}=\frac{4\pi}{m_{n}}$ and
$r=\frac{(m_{\alpha}-m_{n})}{(m_{\alpha}+m_{n})}$. Also,
$\sigma_{2}$ indicates the Pauli matrix so that the spin projection
matrix ($\frac{i}{\sqrt{2}}\sigma_{2}$) projects the two neutrons on
the spin-singlet case. In Eq. (\ref{e4}), the $S_{i}$${}^{,}s$ are
the 2$\times$4 matrices connecting states with total angular
momentum $j=1/2$ and $j=3/2$. These matrices satisfy the following
relations
\begin{eqnarray}
 S_{i}S^{\dag}_{j}&=&\frac{2}{3}\delta_{ij}-\frac{i}{3}\epsilon_{ijk}\sigma_{k},\nonumber\\
S^{\dag}_{j}S_{i}&=&\frac{3}{4}\delta_{ij}-\frac{1}{6}\{J_{i}^{3/2},J_{j}^{3/2}\}+\frac{i}{3}\epsilon_{ijk}J_{k}^{3/2},
\label{e6}
\end{eqnarray}
where $J_{i}^{3/2}$ are the generators of the $J=3/2$ representation
of the rotation group. The parameter $\Delta_{0}$ should be fixed
from matching the pionless EFT $nn$ scattering amplitude to the ERE
scattering amplitude of two non-relativistic nucleons. Also we have
the following relations \cite{h1}
\begin{eqnarray}
g_{1}^{2}=-\eta_{1}\frac{6\pi}{\mu^{2}r_{1}}\;\;,\;\;
\Delta_{1}=\frac{1}{\mu a_{1}r_{1}},
 \label{e7}
\end{eqnarray}
where $\mu$ is the reduced mass of $n\alpha$ system. According to
the sign of $r_{1}$ the sign $\eta_{1}$ should be fixed to +1. Due
to gauge invariance of the non-interacting parts of Lagrangian for
charged alpha and $d_{1}$-dimer, we include electromagnetic coupling
with vector potential $A_{\mu}$. This minimal coupling gives the
covariant derivative as
\begin{eqnarray}
 \partial_{\mu}\rightarrow D_{\mu}=\partial_{\mu}+i\hat{Q}A_{\mu},
\label{e8}
\end{eqnarray}
that $A_{\mu}$ satisfies the Coulomb gauge fixing relation as
$\vec{\nabla}$~$\!\!\cdot\vec{A}=0$. In Eq.~(\ref{e8}), $\hat{Q}$
introduces the charge operator such that $\hat{Q}\phi=\mathcal
{Z}e\phi$, $\hat{Q}d_{1}=\mathcal {Z}ed_{1}$, $\hat{Q}n=0$, and
$\hat{Q}d_{0}=0$, where $\mathcal{Z}$ is the number of protons in
the alpha core.

\section{Two-body and three-body systems}\label{Two-body and three-body systems}
\subsection{Two body propagator}\label{Two body propagator}

\begin{figure}
\centering
\includegraphics[width=8cm]{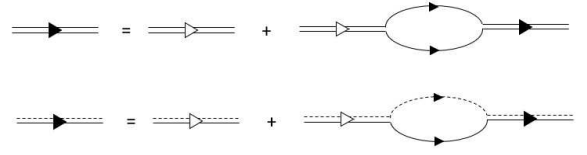}\\
\caption{\small{Dressing the bare $nn\;(n\alpha)$ dimer propagator
with bubble diagrams leads to the full $nn\;(n\alpha)$ dimer
propagator to all orders.  The solid lines show neutron and the
dashed lines are the $\alpha$ particle. Double solid lines with
empty (filled) arrow are the bare (full) $d_{0}$-dimer propagators
and dashed-solid lines with empty (filled) arrow are the bare (full)
$d_{1}$-dimer propagators.}}\label{fig1}
\end{figure}

The full dimer propagators are obtained by the infinite sum of
diagrams shown in Fig. \ref{fig1}. The solid lines indicate neutron
and the dashed lines are the $\alpha$ particle. The bare $nn$-dimer
propagator has been depicted by double solid lines with empty arrow,
and the bare $n\alpha$-dimer propagator is observed by dashed-solid
lines with empty arrow.

Based on introduced power counting, the LO full dimer propagators
shown by filled arrow in Fig. (\ref{fig1}) for auxiliary field
$d_{0}$ and $d_{1}$ are obtained by the following expressions
\begin{eqnarray}
&&iD_{0}(p_{0},\textbf{p})=\frac{i}{\frac{1}{a_{0}}-\sqrt{\frac{\textbf{p}^{2}}{4}-m_{n}p_{0}-i\epsilon}},
\nonumber\\
&&iD_{1}(p_{0},\textbf{p})_{\bar\alpha}^{\bar\beta}=
i\eta_{1}\delta_{\bar\alpha}^{\bar\beta}\bigg\{p_{0}
-\frac{\textbf{p}^{2}}{2(m_{n}+m_{\alpha})}-\frac{1}{\mu a_{1}r_{1}}
\nonumber\\&&\quad-\frac{1}{\mu r_{1}}\big(\frac{\mu}{m_{n}+m_{\alpha}}\textbf{p}^{2}-2\mu p_{0}-i\epsilon\big)^\frac{3}{2}+i\epsilon\bigg\}^{-1},\nonumber\\
\label{e9}
\end{eqnarray}
that the incoming and outgoing spin components of $d_{1}$-dimer is
indicated by $\bar\alpha$ and $\bar\beta$ respectively. Because of
$J=\frac{3}{2}$ of $d_{1}$-dimer, $\delta_{\bar\alpha}^{\bar\beta}$
is a $4\times4$ unit matrix.

\subsection{Three point function}\label{Three point function}
\subsubsection{Full trimer propagator}\label{Full trimer propagator}

\begin{figure}
\centering
\includegraphics[width=8cm]{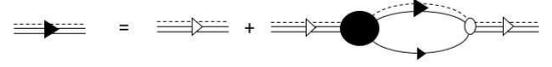}
\caption{\small{Feynman diagrams for the full trimer propagator $t$.
Single, double, triple lines denote neutron, $d_{1}$-dimer and
trimer respectively. Triple line with empty (filled) arrow is the
bare (full) trimer propagator.}}\label{fig2}
\end{figure}

The amplitude of the particle-dimer scattering process in $nn\alpha$
system has been calculated using the Faddeev equation introduced in
Appendix A. The transition amplitude (T-matrix) has a pole at
three-body bound state, so the T-matrix can be factorized at energy
$E=-B_{2n}$ as \cite{h8}
\begin{eqnarray}
T(E,k,p)=
-\frac{\vec{\mathcal{B}}^{\dag}(k)\;\vec{\mathcal{B}}(p)}{E+B_{2n}}\;+\;
\mathrm{regular\;terms}, \label{e26}
\end{eqnarray}
where $B_{2n}=0.97\;\textrm{MeV}$ \cite{h6} denotes the 2n
separation energy of system.
$\vec{\mathcal{B}}(p)={\mathcal{B}_{0}(p)\choose
\mathcal{B}_{1}(p)}$ is the bound state vector, such that
${\mathcal{B}}_{0}\;({\mathcal{B}}_{1})$ corresponds to the $\phi
d_{0}\rightarrow \phi d_{0}\;(\phi d_{0}\rightarrow n d_{1})$
transition of the bound state equation according to Eqs. (\ref{e21})
and (\ref{e22}). Fig. \ref{fig2} indicates the Feynman diagrams
contributing to the full trimer propagator $t(E)$. Based on the
three-body interaction introduced in Eq. (\ref{e20}), the three-body
force appears only between the incoming and outgoing $n+d_{1}$
channels. So, only $T_{11}$ component derived from  $2\times2$
T-matrix integral equation in Eq. (\ref{e12}) contributes to $t(E)$.
Using Feynman rules and taking into account the projection operator
in Eq. (\ref{e11}), for a trimer propagator we can write
\begin{eqnarray}
it(E)&=&\frac{i}{\Omega}\Big[1+\frac{m_{n}\;g_{1}^{2}}{\Lambda^{2}}
\frac{h^{2}}{\Omega}\int_{0}^{\Lambda}dq\;\frac{q^{2}}{2\pi^{2}}\;q^{2}\;
\overline{D}_{1}(E,q)
\nonumber\\
&-&\frac{m_{n}\;g_{1}^{2}}{6\Lambda^{2}}\frac{h^{2}}{\Omega}\int_{0}^{\Lambda}
dq\;\frac{q^{2}}{2\pi^{2}}\;\int_{0}^{\Lambda}dq'\;\frac{q'^{2}}{2\pi^{2}}
\nonumber\\&&\Big(\overline{D}_{1}(E,q')\; q'q \;T_{11}(E,q',q)\;
\overline{D}_{1}(E,q)\Big)\Big], \label{e27}
\end{eqnarray}
where the energy integrals have been carried out and
$\overline{D}_{1}(E,q)=D_{1}(E-\frac{q^{2}}{2m_{n}},q)$.

\subsubsection{Trimer wave function renormalization}\label{Trimer wave function renormalization}
The trimer wave function renormalization constant can be extracted
from the following relation \cite{h8,h39}
\begin{eqnarray}
Z_{t}= \lim_{E\to -E^{(3)}}(E + E^{(3)})\;t(E), \label{e28}
\end{eqnarray}
where $E^{(3)}=B_{2n}$. By neglecting the regular functions in terms
of $E$ corresponding to the first and second terms in Eq.
(\ref{e27}), we substitute the last term of $t(E)$ into Eq.
(\ref{e28}) and finally obtain
\begin{eqnarray}
Z_{t}&=&\frac{m_{n}\;g_{1}^{2}}{6\Lambda^{2}}\frac{h^{2}}{\Omega^{2}}\nonumber\\
&&\lim_{E\to -E^{(3)}}\Big[-(E +
E^{(3)})\int_{0}^{\Lambda}dq\;\frac{q^{3}}{2\pi^{2}}
\;\int_{0}^{\Lambda}dq'\;\frac{q'^{3}}{2\pi^{2}}
\nonumber\\&&\times\Big(\overline{D}_{1}(E,q')\;T_{11}(E,q',q)\;
\overline{D}_{1}(E,q)\Big)\Big]. \label{e29}
\end{eqnarray}
For the incoming and outgoing $n+d_{1}$ channels, inserting Eq.
(\ref{e26}) into Eq. (\ref{e29}) yields

\begin{eqnarray}
Z_{t}&=&\frac{m_{n}\;g_{1}^{2}}{6\Lambda^{2}}\frac{h^{2}}{\Omega^{2}}
\int_{0}^{\Lambda}dq\;\frac{q^{3}}{2\pi^{2}}\;\int_{0}^{\Lambda}dq'\;\frac{q'^{3}}{2\pi^{2}}
\nonumber\\&&\times\Big(\overline{D}_{1}(-B_{2n},q')
\;{\mathcal{B}}^{\dag}_{1}(q')\;{\mathcal{B}}_{1}(q)\; \overline{D}_{1}(-B_{2n},q)\Big)\nonumber\\
&=&\frac{m_{n}\;g_{1}^{2}}{6\Lambda^{2}}\frac{h^{2}}{\Omega^{2}}\;\left|\int_{0}^{\Lambda}dq
\frac{q^{3}}{2\pi^{2}}\overline{D}_{1}(-B_{2n},q){\mathcal{B}}_{1}(q)\right|^{2}.
\label{e30}
\end{eqnarray}

\subsubsection{Trimer-dimer-particle three point function $\vec{\mathcal{G}}^{irr}$}\label{Trimer-dimer-particle three point function}
\begin{figure}
\centering
\includegraphics[width=8.5cm]{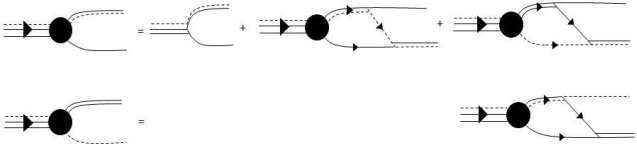}
\caption{\small{Diagrammatic representation of coupled integral
equation for the trimer-dimer-particle three point function
$\vec{\mathcal{G}}^{irr}$.}}\label{fig3}
\end{figure}

The calculation of the charge form factor of ${}^{6}\textrm{He}$
halo nucleus requires the trimer-dimer-particle three point function
$\vec{\mathcal{G}}^{irr}$. The LO $\vec{\mathcal{G}}^{irr}$ function
is illustrated by the coupled integral equation in Fig. \ref{fig3}.
$\vec{\mathcal{G}}^{irr}$ is a vector with two components as
$\vec{\mathcal{G}}^{irr}(E,p)={\mathcal{G}^{irr}_{0}\choose
\mathcal{G}^{irr}_{1}}$, where $\mathcal{G}^{irr}_{0}$
$(\mathcal{G}^{irr}_{1})$ is the three point function for trimer
constructed by $\alpha-d_{0}$ ($n-d_{1}$) system. Using introduced
Lagrangian in Eq. (\ref{e4}) and Feynman rules, we can obtain the
relation for two components of $\vec{\mathcal{G}}^{irr}$ as
\begin{eqnarray}
&&i\;\mathcal{G}_{i}^{irr}(E,k)=i\frac{\sqrt{m_{n}}}{\Lambda}\;g_{1}\;h\;\frac{\sqrt{3}}{2}
\Big[\delta_{1i}\;\vec{S^{\dag}}\cdot \vec{k}\;\sigma_{2}\nonumber\\
&&-\int_{0}^{\Lambda}dq\;\frac{q^{2}}{2\pi^{2}}\;\Big(\vec{S^{\dag}}\cdot
\vec{q}\;\sigma_{2}\Big)\left. \overline{D}_{1}(E,q)\;
T_{1i}(E,q,k)\right|_{H=0}\Big],\nonumber\\ \label{e31}
\end{eqnarray}
where the three-body force $H$ has been introduced in Eq.
(\ref{e20}). All trimer-reducible contributions are neglected by
setting $H=0$ in Eq. (\ref{e31}) \cite{h8}. Taking into
consideration $\left.T_{1i}(E,q,k)\right|_{H=0}$ component of Eq.
(\ref{e12}), we have
\begin{eqnarray}
&&\mathcal{G}_{i}^{irr}(E,k)=\frac{\sqrt{m_{n}}}{\Lambda}\;g_{1}\;h\;\frac{\sqrt{3}}{2}\delta_{1i}\;\vec{S^{\dag}}\cdot \vec{k}\;\sigma_{2}\nonumber\\
&&-\sum_{j=0}^{1}\int_{0}^{\Lambda}dq\frac{q^{2}}{2\pi^{2}}\mathcal{G}_{j}^{irr}(E,q)
\overline{D}_{j}(E,q)\; R_{ji}(E,q,k), \label{ejadid}
\end{eqnarray}
where $i,j=0,1$ denote the components of kernel matrix corresponding
to $R_{00}$, $R_{01}$, $R_{10}$, and $R_{11}$ that have been derived
in Eqs. (\ref{e14})-(\ref{e17}). Taking into account the projection
operator based on Eq. (\ref{e11}), we have
\begin{eqnarray}
\textrm{Tr}\Big(\sqrt{\frac{3}{2}}
\;\sigma_{2}\;(\hat{e_{k}}\cdot\vec{S})\;\vec{S^{\dag}}\cdot
\vec{k}\;\sigma_{2}\Big)=\sqrt{\frac{8}{3}}\;k, \label{ejadid1}
\end{eqnarray}
therefore, the matrix integral equation for the P-wave irreducible
trimer-dimer-particle three point function is given by
\begin{eqnarray}
\vec{\mathcal{G}}^{irr}(E,k)&=&Z_{t}^{\frac{1}{2}}\frac{\sqrt{m_{n}}}{\Lambda}\;g_{1}\;h\;\sqrt{2}\;k\;\hat{\upsilon}\nonumber\\
&-&\int_{0}^{\Lambda}dq\;\frac{q^{2}}{2\pi^{2}}\;\Big(R(E,q,k)\;D(E,q)\;\vec{\mathcal{G}}^{irr}(E,q)\Big),\nonumber\\
\label{e32}
\end{eqnarray}
where $2\times2$ kernel matrix $R$ has been defined in Eq.
(\ref{e14}) and we have
\begin{eqnarray}
 D(E,q)\equiv\left(
                              \begin{array}{cc}
                              \overline{D}_{0}(E,q) & 0  \\
                               0 & \overline{D}_{1}(E,q)\\
\end{array}
\right), \label{e13}
\end{eqnarray}
with $\overline{D}_{0}(E,q)=D_{0}(E-\frac{q^{2}}{2m_{\alpha}},q)$
and $\overline{D}_{1}(E,q)=D_{1}(E-\frac{q^{2}}{2m_{n}},q)$.

Two components of Eq. (\ref{e32}) that enter into the calculation of
form factor for the ${}^{6}\textrm{He}$ halo nucleus are derived as
\begin{eqnarray}
\mathcal{G}_{i}^{irr}(E,k)&=&\frac{1}{\sqrt{3}}\;\frac{m_{n}\;g_{1}^{2}}{\Lambda^{2}}\;\left|\beta\;H_{0}(\Lambda)\right|\Big[k\;\delta_{1i}\nonumber\\
&-&\int_{0}^{\Lambda}dq\;\frac{q^{2}}{2\pi^{2}}\;\Big(q\;\left.
\overline{D}_{1}(E,q)\;
T_{1i}(E,q,k)\right|_{H=0}\Big)\Big],\nonumber\\ \label{e33}
\end{eqnarray}
where
\begin{eqnarray}
\beta=\int_{0}^{\Lambda}dq\;\frac{q^{2}}{2\pi^{2}}\;q\;\overline{D}_{1}(-B_{2n},q)\;{\mathcal{B}}_{1}(q),
\label{e34}
\end{eqnarray}
and the calculation of $H_{0}(\Lambda)=\frac{h^{2}}{\Omega}$
requires the normalized ${}^{6}\textrm{He}$ wave function which is
obtained by solving the bound state equation corresponding to the
homogeneous part of Eq. (\ref{e12}) with $E=-B_{2n}$. We should
mention that $(\hat{\upsilon})_{i}=\delta_{1i}$ and after
multiplying $\sqrt{Z_{t}}$ in Eq. (\ref{e32})
$\vec{\mathcal{G}}^{irr}$ has the cutoff dependence which is small
enough to render our predictions renormalized.

\section{Charge form factor of ${}^{6}\textrm{He}$ halo nucleus}\label{Charge form factor}

We present a formalism for form factor calculation of
${}^{6}\textrm{He}$ halo nucleus with shallow P-wave interaction. We
initially emphasize that all calculations have been performed in the
Breit frame in which no energy is carried by the photon. This
implies $P_{0}=K_{0}$ and $\vec{P}^{2}=\vec{K}^{2}$ where $\vec{P}$
$(\vec{K})$ denotes the incoming (outgoing) three momentum of
trimer. The ${}^{6}\textrm{He}$ charge form factor only depends on
the three-momentum of the photon $\vec{Q}^{2}=(\vec{K}-\vec{P})^{2}$
according to \cite{h9,h40}
\begin{eqnarray}
\langle t(K_{0}, \vec{K})|j_{0}|t(P_{0},
\vec{P})\rangle=(-ie\mathcal{Z})\mathcal{F}_{E}(\vec{Q}^{2})=Z_{t}i\Gamma(\vec{Q}),\nonumber\\
\label{e35}
\end{eqnarray}
where $\mathcal{Z}$ implies the atomic number of ${}^{4}\textrm{He}$
nucleus and $j_{0}$ is the zeroth component of the electromagnetic
current.

\begin{figure}
\centering
\includegraphics[width=8cm]{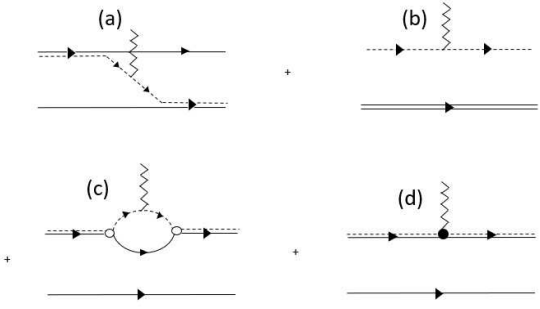}
\caption{\small{Diagrammatic representation of the matrix element
$\overline{\Gamma}(\vec{Q})$ in Eq. (\ref{e36})} falling into four
different classes (a), (b), (c) and (d). The wavy lines show
minimally photon coupled and the vertex depicted by filled circle in
diagram (d) indicates minimally coupled photon to the P-wave
$d_{1}$-dimer.}\label{fig4}
\end{figure}
\begin{figure}
\centering
\includegraphics[width=8cm]{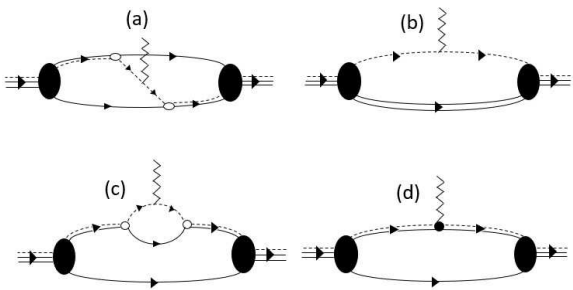}
\caption{\small{Diagrams for the charge form factor of the
${}^{6}\textrm{He}$ nucleus at LO corresponding to
$\mathcal{F}_{E}(\vec{Q}^{2})$ in Eq.(\ref{e35}). Notations are as
in Fig. (\ref{fig4}). }}\label{fig5}
\end{figure}

At LO, $\Gamma(\vec{Q})$ in Eq. (\ref{e35}) introduces the sum of
all diagrams with external trimer lines and minimally coupled photon
to the alpha and $d_{1}$-dimer as shown in Fig. \ref{fig5}, so we
have
\begin{eqnarray}
i\Gamma(\vec{Q})&=&\int\;\frac{d^{4}p}{(2\pi)^{4}}\int\;\frac{d^{4}k}{(2\pi)^{4}}\nonumber\\
&\times&i\vec{\mathcal{G}}^{irr}(E, \vec{P}, p_{0},
\vec{p})^{T}\;i\overline{\Gamma}(E, \vec{P}, p_{0}, \vec{p}, \vec{K}, k_{0},
\vec{k})\; \nonumber\\&\times&i \vec{\mathcal{G}}^{irr}(E, \vec{K},
k_{0}, \vec{k}). \label{e36}
\end{eqnarray}

The matrix element $\overline{\Gamma}$ in Eq. (\ref{e36}) is defined
by the sum of diagrams that are depicted in Fig. \ref{fig4}. The
energy quantity is defined by $E= P_{0}-\frac{\vec{P}^2}{2
M_{tot}}$, where the kinetic energy of the ${}^{6}\textrm{He}$
system is subtracted and $M_{tot}= m_{\alpha}+2m_{n}$. Therefore,
the ${}^{6}\textrm{He}$ charge form factor at LO  is found by the
sum of diagrams in Fig. \ref{fig4} as
$\mathcal{F}_{E}=\mathcal{F}_{E}^{(a)}+\mathcal{F}_{E}^{(b)}+\mathcal{F}_{E}^{(c)}+\mathcal{F}_{E}^{(d)}$.
The wavy lines show minimally coupled photon and the vertex depicted
by filled circle in diagram (d) indicates minimally coupled photon
to the P-wave $d_{1}$-dimer.\\ For calculating of the charge form factor, it is necessary to define the relation
between $\mathcal{G}_{i}^{irr}(E, \vec{P}, p_{0}, \vec{p})$ and
center-of-mass $(c.m.)$ quantity $\mathcal{G}_{i}^{irr}(E,p)$ from
Eq. (\ref{e33}) via the following integral equation
\begin{eqnarray}
&&\mathcal{G}_{i}^{irr}(E, \vec{P}, p_{0}, \vec{p})=\frac{1}{\sqrt{3}}\;\frac{m_{n}\;g_{1}^{2}}{\Lambda^{2}}\;\left|\beta\;H_{0}(\Lambda)\right|p\;\delta_{1i}\nonumber\\
&-&\;\sum_{j=0}^{1}\int_{0}^{\Lambda}dq\frac{q^{2}}{2\pi^{2}}\;R_{ij}\Big(\frac{M_{i}}{M_{tot}}E+p_{0}-\frac{\vec{P}\cdot\vec{p}}{M_{tot}}+\frac{p^{2}}{2m_{i}},p,q\Big)
\nonumber\\&&\qquad\times\overline{D}_{j}(E,q)\;\mathcal{G}_{j}^{irr}(E,q),\nonumber\\
\label{a1}
\end{eqnarray}
where $m_{i}=m_{n},m_{\alpha}$ for $i=0,1$, respectively, and
\begin{eqnarray}
M_{0}=2m_{n},\;\;\;M_{1}=m_{\alpha}+m_{n}. \label{a2}
\end{eqnarray}

After performing calculations according to Eq.~(\ref{e4}) and using
Feynman rules, we obtain the following final relations for charge
form factor contributions of diagrams (a), (b), (c) and (d) in Fig.
5
 \begin{eqnarray}
\mathcal{F}^{(a)}(Q^{2})&=&\int_{0}^{\Lambda}
\frac{p^{2}}{2\pi^{2}}\;dp\;\int_{0}^{\Lambda}
\frac{k^{2}}{2\pi^{2}}\;dk\;\vec{\mathcal{G}}^{irr}(p)^{T}
\nonumber\\
&&\times D(p)\;\Upsilon^{(a)}(Q,p,k)
D(k)\;\vec{\mathcal{G}}^{irr}(k), \label{e70}
\end{eqnarray}

\begin{eqnarray}
\mathcal{F}^{(b)}(Q^{2})&=&\;\int_{0}^{\Lambda}
\Big(-\frac{p^{2}}{2\pi^{2}}\Big)\;dp\;\int_{0}^{\Lambda}
\Big(-\frac{k^{2}}{2\pi^{2}}\Big)\;dk
\nonumber\\&\times&\vec{\mathcal{G}}^{irr}(p)^{T}\;D(p)
\;\Upsilon^{(b)}(Q,p,k)\;D(k)\;\vec{\mathcal{G}}^{irr}(k),\nonumber\\
\label{e71}
\end{eqnarray}

\begin{eqnarray}
&&\mathcal{F}^{(c)}(Q^{2})=\;\int_{0}^{\Lambda} \Big(-\frac{p^{2}}{2\pi^{2}}\Big)\;dp\;\int_{0}^{\Lambda}
\Big(-\frac{k^{2}}{2\pi^{2}}\Big)\;dk\;\vec{\mathcal{G}}^{irr}(p)^{T}
\nonumber\\&&\times D(p)\;\Upsilon^{(c)}(Q,p,k)\;D(k)\;\vec{\mathcal{G}}^{irr}(k)\nonumber\\
&&+2\int_{0}^{\Lambda}
\Big(-\frac{p^{2}}{2\pi^{2}}\Big)dp\vec{\mathcal{G}}^{irr}(p)^{T}D(p)\vec{\Upsilon}^{(c)}(Q,p)+\Upsilon^{(c)}_{0}(Q),\nonumber\\
\label{e72}
\end{eqnarray}
and
\begin{eqnarray}
\mathcal{F}^{(d)}(Q^{2})&=&\;\int_{0}^{\Lambda} \Big(-\frac{p^{2}}{2\pi^{2}}\Big)\;dp\;\int_{0}^{\Lambda}
\Big(-\frac{k^{2}}{2\pi^{2}}\Big)\;dk
\nonumber\\&&\quad\times\vec{\mathcal{G}}^{irr}(p)^{T}\;D(p)\;\Upsilon^{(d)}(Q,p,k)\;D(k)\;\vec{\mathcal{G}}^{irr}(k)\nonumber\\
&&+2\int_{0}^{\Lambda}
\Big(-\frac{p^{2}}{2\pi^{2}}\Big)\;dp\;\vec{\mathcal{G}}^{irr}(p)^{T}\;D(p)\;\vec{\Upsilon}^{(d)}(Q,p)
\nonumber\\&&+\Upsilon^{(d)}_{0}(Q). \label{e73}
\end{eqnarray}
The detailed derivations of Eqs. (\ref{e70})-(\ref{e73}) including
the definitions of the used functions have been explained in
Appendix~\ref{Appendix B}.

\section{Numerical calculation and results}\label{Numerical calculation and results}
As we know at $Q^{2}=0$, the charge form factor is normalized to one
because of conservation of current. The expansion of the form factor
in powers of $Q^{2}$ leads to
\begin{eqnarray}
\mathcal{F}_{E}(Q^{2})=\;1-\;\frac{\langle
r_{E}^{2}\rangle}{6}\;Q^{2}+ \cdots, \label{e37}
\end{eqnarray}
where $\langle r_{E}^{2}\rangle$ is the mean-square charge radius of the
${}^{6}\textrm{He}$ halo system relative to ${}^{4}\textrm{He}$
mean-square charge radius. By taking the limit $Q^{2}\rightarrow0^{+}$, $\langle
r_{E}^{2}\rangle$ can be extracted as
\begin{eqnarray}
\langle r_{E}^{2}\rangle=\;-6\lim_{Q^{2}\to
0^{+}}\frac{d\mathcal{F}_{E}}{dQ^{2}}, \label{e38}
\end{eqnarray}
and we can obtain the mean-square charge radius of ${}^{6}\textrm{He}$ halo
nucleus by the following relation
\begin{eqnarray}
\langle r_{E}^{2}\rangle_{{}^{6}He}=\langle
r_{E}^{2}\rangle\;+\langle r_{E}^{2}\rangle_{{}^{4}He}. \label{e39}
\end{eqnarray}

It is necessary to point out that we have neglected the small
negative mean-square charge radius of the neutron $\langle
r_{E}^{2}\rangle_{n}=-0.115\;\textrm{fm}^{2}$ \cite{h10} in our
calculation. In this section, we apply our P-wave halo EFT formalism
to calculate the form factor and the mean-square charge radius of
${}^{6}\textrm{He}$ nucleus relative to ${}^{4}\textrm{He}$ core
$(\langle r_{E}^{2}\rangle)$ according to Eq. (\ref{e38}). We
compare our EFT evaluation with other available theoretical results.
Our formalism applies directly to two-neutron halo nucleus,
${}^{6}\textrm{He}$ with $J^{P}=0^{+}$.

\begin{figure*}[tb]\centering
\includegraphics*[width=15cm]{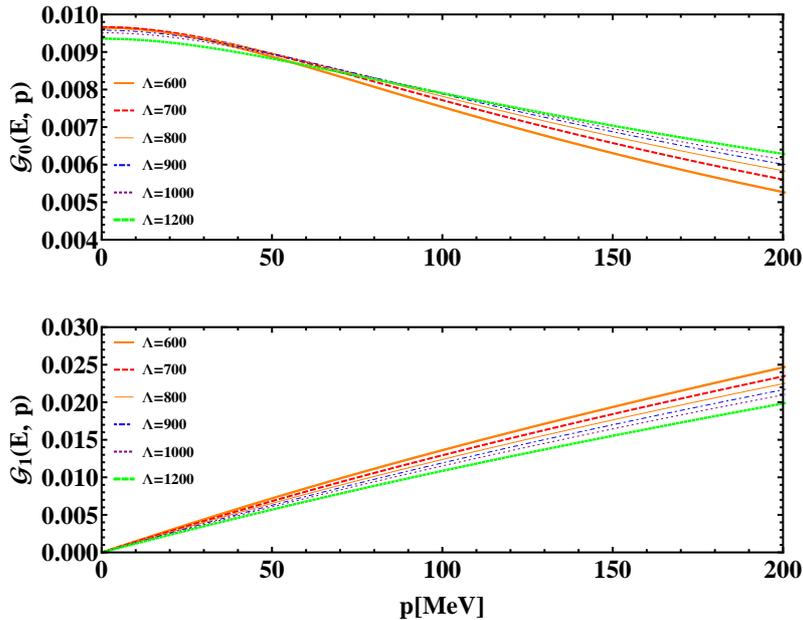} \caption{\small{(color online) Two components of the
trimer-dimer-particle three point function for the different cutoff
values from 600 MeV to 1200 MeV. }}\label{fig10}
\end{figure*}

Fitting the three-body binding energy of ${}^{6}\textrm{He}$ nucleus
to $B_{2n}=0.97 \;\textrm{MeV}$, the three-body force can be
determined at leading order. Using this determined three-body force
\cite{h3}, the renormalized ${}^{6}\textrm{He}$ wave function and so
the renormalized trimer-dimer-particle three point function is
obtained. The effects of the cutoff dependence for the two
components of $\vec{\mathcal{G}}$ function are shown in
Fig.\ref{fig10}. The plots represent the cutoff variations of the
P-wave irreducible trimer-dimer-particle three point function
between $\Lambda=600$ MeV and $\Lambda=1200$ MeV with the three-body
force which is introduced by Eq.~(\ref{e20}). As depicted in
Fig.\ref{fig10}, by considering the three-body force, the cutoff
variations of the results are acceptable in comparison with the LO
systematical uncertainty. Therefore our numerical results for the
trimer-dimer-particle three point function are properly
renormalized.

As mentioned in Section \ref{Power-counting}, we concentrate on the
power counting which is suggested by Bertulani and collaborators in
Ref.~\cite{h1} for the $n\alpha$ interaction. Investigation of the
full propagator of the $d_1$ field in this power counting discloses
in addition to the physical resonance (shallow na resonance), there
also exists one spurious pole (the unphysical ${}^{5}\textrm{He}$
bound state) around $p\sim99$ MeV with negative residue and a deeper
binding energy. Using this power counting, the Gamow shell model
calculation of ${}^{6}\textrm{He}$ in Ref.~\cite{h5} removed the
spurious pole in the $n\alpha$ T-matrix by constructing
bi-orthogonal complete basis.

\begin{center}
\begin{table}
\centering {\small{\caption{\label{t1} The mean-square charge radius of
${}^{6}\textrm{He}$ nucleus relative to ${}^{4}\textrm{He}$ from Eq.
(\ref{e38}) and the r.m.s charge radius of ${}^{6}\textrm{He}$ nucleus
according to Eq. (\ref{e39}) that have been compared to the other
theoretical and experimental results. }}}
\begin{tabular}{c|c|c|c}
\hline \hline
                                                        &  This     &   Experimental    &  Theoretical  \\
                                                        &  work     &   Results         &  Results  \\
\hline
  $\langle r_{E}^{2}\rangle[\textrm{fm}^{2}]$                    & 1.408      &    $1.418 \pm 0.058 $ \cite{h31} &  1.426(38) \cite{h35} \\
                                                        &                &    $1.047\pm0.034 $ \cite{h32}   &                       \\
\hline
                                                                   &           &  2.060(8) \cite{h33}          &  2.06(1) \cite{h35}     \\
   $\langle r_{E}^{2}\rangle_{{}^{6}He}^{\frac{1}{2}}[\textrm{fm}]$&  $2.058$ & $2.054\pm0.014$ \cite{h31}    &  2.12(1) \cite{h35}      \\
                                                                   &           &                               &  2.147 \cite{h29}    \\
                                                                   &           &                               & 2.586 \cite{h34} \\
\hline \hline
 \end{tabular}
\end{table}
\end{center}
\begin{table*}[tb]\centering
\caption{\label{t2} The mean-square charge radius of ${}^{6}\textrm{He}$ nucleus
relative to ${}^{4}\textrm{He}$ from Eq. (\ref{e38}) and the r.m.s
charge radius of ${}^{6}\textrm{He}$ nucleus according to Eq.
(\ref{e39}) for different cutoff values. }
\begin{tabular}{|c||c|c|c|c|c|c|}
\hline
                    $\Lambda$ (MeV)                                    & 600   & 700  & 800   & 900 & 1000 & 1200   \\
\hline\hline
  $\langle r_{E}^{2}\rangle[\textrm{fm}^{2}]$                    & $\quad$1.40822$\quad$  & $\quad$1.40807$\quad$& $\quad$1.40739$\quad$& $\quad$1.40786$\quad$ & $\quad$1.40801$\quad$ & $\quad$1.40777$\quad$  \\
\hline $\quad\langle
r_{E}^{2}\rangle_{{}^{6}He}^{\frac{1}{2}}[\textrm{fm}]\quad$    & 2.05766  & 2.05763& 2.05746& 2.05758& 2.05761& 2.05755  \\
\hline
 \end{tabular}
\end{table*}

In the Halo EFT analysis of the ${}^{6}\textrm{He}$ system, in order
to get rid of this spurious pole, one can also treat the unitarity
term $ik^3$ in the denominator of $n\alpha$ propagator as a
perturbation. This method was recently applied to
${}^{6}\textrm{He}$ in Ref.~\cite{h4,Ryberg}. One of fundamental
drawbacks of this method is that unitarity is lost at LO, which is
actually a requirement for the form $\frac{1}{p\cot\delta_l-ip}$
that was chosen for the scattering amplitude. In fact the loss of
unitarity at LO is not problematic in Ref.~\cite{h4}, since only
bound state observable is considered in the related bound-state
three-body calculation. In the Faddeev equation, the resonance pole
of $n\alpha$ scattering, which requires the unitary term, was never
crossed. However, the unitary term matters if one wants to calculate
a resonance state in ${}^{6}\textrm{He}$.

Generally, in the three-body sector, for solving Faddeev integral
equation, analogous to the Skornyakov-Ter-Martyrosian (STM) equation
for S-wave contact interactions, one can solve Faddeev integral
equation for resonant P-wave interactions. In order to discard
spurious pole one can use the contour deformation suggested by
Hetherington and Schick, namely a rotation $p\rightarrow
pe^{-i\Phi}$ ($\Phi>0$) as applied in Ref.~\cite{h3} for the
positive energies.

In this paper we are concerned with the homogeneous part of the
integral equations projected onto the bound 0+ ground state of
${}^{6}\textrm{He}$. The position of spurious bound state of a
P-wave propagator is on the real axes but for the negative energies
$E=E_{B}=-B_{2n}$, Eq.~(\ref{e21}). One can handle this unphysical deep
bound state with similar $p\rightarrow pe^{-i\Phi}$ ($\Phi>0$)
analytical continuation by contour rotation of the real axes. In
this simpler contour path integral, there is no logarithmic
singularities in the loop momentum in comparison with logarithmic
singularities in the Legendre functions of second kind in the
positive energies on the real axes.

\begin{figure}
\centering
\includegraphics[width=8cm]{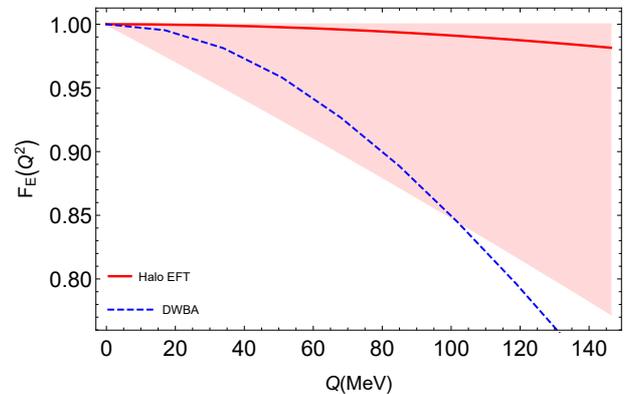} \caption{\small{(color online) The charge form factor of the
${}^{6}\textrm{He}$ nucleus with $\Lambda=700$ MeV at LO
corresponding to $\mathcal{F}_{E}(\vec{Q}^{2})$ in Eq. (\ref{e35}).
The shaded region implies a criterion of estimated theoretical
artifacts.}}\label{fig8}
\end{figure}

In Fig. \ref{fig8}, our calculation for the charge form factor of
${}^{6}\textrm{He}$ with $\Lambda=700$ MeV is depicted as a function
of the photon momentum $(\vec{Q}=\vec{K}-\vec{P})$ in halo EFT. Our
results have been compared with distorted wave Born approximation
(DWBA) \cite{h29} large scale shell model (LSSM) calculations. This
model-dependent approach studies ${}^{6}\textrm{He}$ nucleus as a
six-body system (not three-body halo one) by using a Woods-Saxon
single-particle wave function basis. The difference in the
calculated form factors appears as the $Q$ is increasing. As we
expect EFT is a model-independent and precision-controlled approach,
so including the higher-order corrections can give us better
judgment in comparing our three-particle halo formalism with the
full six-body DWBA results.

Using Eq. (\ref{e38}), we fit our data for form factor via the
standard interpolation method and take the second order derivatives
of the fitted function with respect to Q in the limit
$Q^{2}\rightarrow0^{+}$ to evaluate the mean-square charge radius of
${}^{6}\textrm{He}$ nucleus relative to ${}^{4}\textrm{He}$ core.
Ingo Sick has studied the world data on elastic electron-Helium
scattering to determine a precise value for ${}^{4}\textrm{He}$
r.m.s charge radius and has obtained the value
$(1.681\pm0.004)\;\textrm{fm}$ \cite{h30}. In Table~\ref{t1}, we
have summarized our EFT evaluated values for the mean-square charge
radius of ${}^{6}\textrm{He}$ nucleus relative to
${}^{4}\textrm{He}$ from Eq.~(\ref{e38}) and the r.m.s charge radius
of ${}^{6}\textrm{He}$ nucleus in comparison with other theoretical
and experimental results.

As discussed in the introduction, the expansion parameter of our
theory $R_{core}/R_{halo}$ is roughly $R/a$. In order to obtain
better estimates, we compare the typical energy scales $E_{halo}$
and $E_{core}$ of the two-neutron halo and the core, respectively.
To estimate $E_{halo}$, we choose the two-neutron separation energy
$B_{2n}$. The energy scale of the core is estimated by the
excitation energy of the alpha particle $E_\alpha$. The square root
of the energy ratio $R_{core}/R_{halo}\sim \sqrt{E_{halo}/E_{core}}$
then yields an estimate for the expansion parameter of the effective
theory. The two-neutron separation energy of ${}^{6}\textrm{He}$ is
0.97 MeV and the first excitation energy of the alpha particle is
20.21 MeV. The expansion parameter and the error can be estimated as
$R_{core}/R_{halo}\sim \sqrt{B_{2n}/E_\alpha}\sim0.22$. So, the
calculated mean-square charge radius and the r.m.s charge radius of
${}^{6}\textrm{He}$ nucleus in Table~\ref{t1} have LO systematic
errors of order of $\delta\langle
r_{E}^{2}\rangle=0.308\,\textrm{fm}^{2}$ and $\delta\langle
r_{E}^{2}\rangle^{\frac{1}{2}}_{{}^{6}He}=0.555\,\textrm{fm}$,
respectively. The expansion parameter $R_{core}/R_{halo}$ is
typically not much smaller than 1. As a consequence, the main
uncertainty in our calculation is from the next-to-leading order
corrections in the effective theory.

The small and negligible cutoff variation in the calculated values
of the r.m.s charge radius of ${}^{6}\textrm{He}$ nucleus as
presented in Table~\ref{t2} shows that our EFT results have been
properly renormalized. The r.m.s charge radius of
${}^{6}\textrm{He}$ nucleus has been determined to a precision of
$0.7$ percent $(2.054\pm0.014)\;\textrm{fm}$ and a difference
between the mean-square charge radii $\langle
r_{c}^{2}\rangle_{{}^{6}He}-\langle r_{c}^{2}\rangle_{{}^{4}He}$ has
been evaluated $(1.418 \pm 0.058)\;\textrm{fm}^{2}$ in a laser
spectroscopic measurement at Argonne National Laboratory \cite{h31}.
Isotope shifts of the matter radii have been deduced via scattering
of GeV/nucleon nuclei on Hydrogen in inverse kinematics. This
approach leads to the value $(1.047\pm0.034)\;\textrm{fm}^{2}$ for
the mean-square charge radius of ${}^{6}\textrm{He}$ isotope
relative to ${}^{4}\textrm{He}$ \cite{h32}. The first direct mass
measurement of ${}^{6}\textrm{He}$ has been performed with the TITAN
Penning trap mass spectrometer at the ISAC facility \cite{h33}. The
obtained mass is $m({}^{6}\textrm{He})=6.018885883(57) u$. With this
new mass value and the previously measured atomic isotope shifts,
they have obtained the r.m.s charge radii of $2.060(8)\;\textrm{fm}$
for ${}^{6}\textrm{He}$ \cite{h33}. Antonov et al. have also
calculated the value $2.147\; \textrm{fm}$ for the r.m.s charge
radius of ${}^{6}\textrm{He}$ nucleus using LSSM densities
\cite{h29}. R.m.s radius in fm for ${}^{6}\textrm{He}$ has
calculated $2.586\; \textrm{fm}$ using the shell model wave
functions and the specified single particle wave functions
\cite{h34}. Our results are consistent to the Monte Carlo
calculation based on AV18+IL2 three-body potential that reports
$2.06(1)\; \textrm{fm}$ for the r.m.s charge radius of
${}^{6}\textrm{He}$ \cite{h35}. Using AV18+UIX three-body potential,
the r.m.s charge radius of ${}^{6}\textrm{He}$ has been obtained
$2.12(1)\;\textrm{fm}$ \cite{h35}.

\section{Conclusion}\label{Conclusion}

In the present halo EFT formalism, we have described the
electromagnetic structure of ${}^{6}\textrm{He}$ halo nucleus. The
trimer propagator and the trimer wave function renormalization
$(Z_{t})$ are obtained in details. The trimer-dimer-particle three
point function $\vec{\mathcal{G}}^{irr}$ that is required for
calculations of form factor is discussed completely. The main
purposes of the present work are the calculation of the charge form
factor and the r.m.s charge radius of ${}^{6}\textrm{He}$. The charge
form factor of ${}^{6}\textrm{He}$ has been obtained by the
summation of four different diagrams depicted in Fig. \ref{fig5}. We
have presented our EFT results for form factor in Fig. \ref{fig8}
and we have shown the shaded region that implies a criterion of
estimated theoretical artifacts in our calculations. The mean-square charge
radius of ${}^{6}\textrm{He}$ nucleus relative to
${}^{4}\textrm{He}$ core and the r.m.s charge radius of
${}^{6}\textrm{He}$ nucleus have been evaluated as $\langle
r_{E}^{2}\rangle=1.408\; \textrm{fm}^{2}$ and $\langle
r_{E}^{2}\rangle_{{}^{6}He}^{\frac{1}{2}}=2.058\; \textrm{fm}$,
respectively with remarkable agreement with other experimental and
theoretical results. In the future works, this formalism can be
expanded to next-to-leading order (NLO) in order to reduce EFT
theoretical error. The ${}^{17}\textrm{B}$ nucleus can be also
described using this P-wave halo EFT approach in the future.

\section*{Acknowledgement}
We would like to thank A. N. Antonov and M. K. Gaidarov for
providing the data of their DWBA calculation.

\appendix
\numberwithin{equation}{section} \makeatletter
\newcommand{\section@cntformat}{Appendix \thesection:\ }
\makeatother

\section{The Faddeev equation of the  particle-dimer scattering process in $nn\alpha$ system} \label{Appendix A}
\begin{figure}
\centering
\includegraphics[width=9cm]{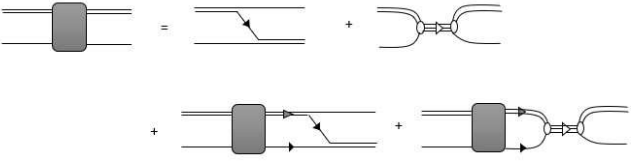}
\caption{\small{Representation of the integral equation for the
T-matrix. Single, double, triple lines denote particles, dimers and
trimers respectively. Triple lines with empty arrow are the bare
trimer propagators.}}\label{fig9}
\end{figure}

Since ${}^{6}\textrm{He}$ nucleus has spin-parity $J^{P}=0^{+}$ in
the ground state, we apply the Faddeev equation (T-matrix) of the
particle-dimer scattering process in $nn\alpha$ system with
$J^{P}=0^{+}$. This integral equation is shown in Fig. \ref{fig9}.
According to Lagrangian in Eq. (\ref{e4}) we use two different
dimers, $d_{0}$ and $d_{1}$, so there are four possible transitions
between particle-dimer states
\begin{eqnarray}
 n+d_{1}&\longrightarrow& n+d_{1},\;\;\;\;\;\;\;\; n+d_{1}\longrightarrow \phi+d_{0},\nonumber\\
\phi+d_{0}&\longrightarrow&
n+d_{1},\;\;\;\;\;\;\;\phi+d_{0}\longrightarrow\phi+d_{0}.
\label{e10}
\end{eqnarray}
In the $c.m.$ frame, on-shell T-matrix depends on the total energy
$E$ and the incoming (outgoing) three-momentums of the $\phi+d_{0}$
and $n+d_{1}$ systems which indicated by $k_{1}\;(p_{1})$ and
$k_{2}\;(p_{2})$ respectively.

In the cluster-configuration space, the projection operator of
${}^{6}\textrm{He}$ channel is obtained by
\begin{eqnarray}
 (\mathcal{P}_{0^{+}}){}^{\beta_{1},\beta_{2}}_{\bar{\alpha},\alpha}=\left(
                              \begin{array}{cc}
                               \bf{1}{}^{\beta_{1}}_{\alpha} & 0 \\
                             0 & \sqrt{\frac{3}{2}}\Big(\sigma_{2}\hat{e}\cdot\vec{S}\Big){}^{\beta_{2}}_{\bar{\alpha}}\\
\end{array}
\right), \label{e11}
\end{eqnarray}
where $\bf{1}$ is the $2\times2$ unit matrix and $\hat{e}$ denotes
the unit vector of $c.m.$ momentum of the $n+d_{1}$ system
\cite{h3}. Applying the projection operator according to Eq.
(\ref{e11}), the resulting  $2\times2$ T-matrix integral equation
can be given by
\begin{eqnarray}
&&T(E,k_{1},p_{1},k_{2},p_{2})=\Big[R(E,k_{1},p_{1},k_{2},p_{2})+ H(k_{2},p_{2},\Lambda)\Big]\nonumber\\
&&\qquad-\frac{1}{2\pi^{2}}\int_{0}^{\Lambda}q^{2}dq\Big[\Big(R(E,k_{1},q,k_{2},q)+ H(k_{2},q,\Lambda)\Big)\nonumber\\
&&\qquad\qquad\qquad\qquad\quad\cdot D(E,q)\cdot
T(E,q,p_{1},q,p_{2})\Big],\label{e12}
\end{eqnarray}where $\Lambda$ is an ultraviolet cutoff.
The kernel $R(E,k_{1},q_{1},k_{2},q_{2})$ is a $2\times2$ matrix
introduced by \cite{h3}
\begin{eqnarray}
 R(E,k_{1},q_{1},k_{2},q_{2})\equiv\left(
                              \begin{array}{cc}
                               0 & R_{10}(E,k_{2},q_{1}) \\
                              R_{01}(E,k_{1},q_{2}) & R_{11}(E,k_{2},q_{2}) \\
\end{array}
\right),\nonumber\\ \label{e14}
\end{eqnarray}
that
\begin{eqnarray}
&&R_{11}(E,k,q)=-\frac{g_{1}^{2}m_{\alpha}}{6}\Big[2(1-r)\frac{k^{2}
+q^{2}}{k q}Q_{1}(\varepsilon_{11}(E,k,q))\nonumber\\
\nonumber\\&&\quad+\frac{8}{3}Q_{2}(\varepsilon_{11}(E,k,q))+(\frac{4}{3}+(1-r)^{2})Q_{0}(\varepsilon_{11}(E,k,q))\Big],\nonumber\\
\label{e15}
\end{eqnarray}
\begin{eqnarray}
R_{10}(E,k,q)&=&-\frac{ g_{0}
g_{1}m_{n}}{4\sqrt{3}}\Big[\frac{2}{k}Q_{1}(\varepsilon_{10}(E,k,q))
\nonumber\\
&&\qquad\quad+\frac{1+r}{q}Q_{0}(\varepsilon_{10}(E,k,q))\Big],
\label{e16}
\end{eqnarray}
\begin{eqnarray}
R_{01}(E,k,q)&=&-\frac{g_{0}
g_{1}m_{n}}{4\sqrt{3}}\Big[\frac{2}{q}Q_{1}(\varepsilon_{01}(E,k,q))
\nonumber\\
&&\qquad\quad+\frac{1+r}{k}Q_{0}(\varepsilon_{01}(E,k,q))\Big].
\label{e17}
\end{eqnarray}
The relation between the $L-th$ Legendre function of the first kind
$P_{L}(z)$ and the second kind $Q_{L}(z)$ is written by
$Q_{L}(z)=\frac{1}{2}\int_{-1}^{1}dt\frac{P_{L}(t)}{z-t}$, therefore
\begin{eqnarray}
Q_{0}(z)&=&\frac{1}{2}\mathrm{ln}\Big(\frac{z+1}{z-1}\Big),\nonumber\\
Q_{1}(z)&=&\frac{1}{2}z \mathrm{ln}\Big(\frac{z+1}{z-1}\Big)-1,\nonumber\\
Q_{0}(z)&=&-\frac{3}{2}z+\frac{1}{4}(3z^{2}-1)\mathrm{ln}\Big(\frac{z+1}{z-1}\Big).
\label{e18}
\end{eqnarray}
The functions of $\varepsilon_{11}$, $\varepsilon_{10}$, and
$\varepsilon_{01}$ in the above equation are defined by \cite{h3}
\begin{eqnarray}
\varepsilon_{11}(E,k,q)&=&\frac{m_{\alpha}\;E - \frac{m_{\alpha}}{2\mu}(k^{2}+q^{2})}{k q},\nonumber\\
\varepsilon_{10}(E,k,q)&=&\frac{m_{n}\;E - k^{2} - \frac{m_{n}}{2\mu}q^{2}}{k q},\nonumber\\
\varepsilon_{01}(E,k,q)&=&\frac{m_{n}\;E - \frac{m_{n}}{2\mu}k^{2} -
q^{2}}{k q}. \label{e19}
\end{eqnarray}
In Eq. (\ref{e12}), the three-body force $H$ shown by a bare trimer
with external particle-dimer lines in Fig. \ref{fig9} is given by
the following relation
\begin{eqnarray}
H(k,q,\Lambda)\equiv\left(
                              \begin{array}{cc}
                               0 & 0 \\
                              0 & -\frac{m_{n}\;g_{1}^{2} k q H_{0}(\Lambda)}{\Lambda^{2}}\\
\end{array}
\right), \label{e20}
\end{eqnarray}
which connects only the incoming and outgoing $n+d_{1}$ channels
\cite{h4,h3,h5}. The bound state equation is written as
\begin{eqnarray}
T_{{}^{6}He}(p)&=&\!-\!\frac{1}{2\pi^{2}}\int_{0}^{\Lambda}q^{2}dq
\Big[R(-B_{2n},k,q,k,q)\!+\! H(k,q,\Lambda)\Big]
\nonumber\\
&&\qquad\cdot D(-B_{2n},q)\cdot T_{{}^{6}He}(q),\label{e21}
\end{eqnarray}
The transition $xX\rightarrow yY$ $(x,y = \phi, n$ and $X, Y =
d_{0}, d_{1})$ contributes to construction of ${}^{6}\textrm{He}$
such that
\begin{eqnarray}
 \!\!\!\!\!T_{{}^{6}He}(q)\!\equiv\!\left(\!
                              \begin{array}{cc}
                               T_{{}^{6}He,\;\phi d_{0}\rightarrow \phi d_{0}}(q) & T_{{}^{6}He,\;n d_{1}\rightarrow\phi d_{0}}(q) \\
                               T_{{}^{6}He,\;\phi d_{0}\rightarrow n d_{1}}(q) & T_{{}^{6}He,\;n d_{1}\rightarrow n d_{1}}(q)\\
\end{array}
\!\right). \label{e22}
\end{eqnarray}
For the incoming  $\phi+d_{0}$ channel, the proper normalization
condition for the solution of Eq. (\ref{e21}) is \cite{h7}
\begin{eqnarray}
\Big(D\vec{\mathcal{B}}\Big)^{T}\otimes
\left.\frac{d}{dE}\Big(I-K\Big)\right|_{E=-B_{2n}}\otimes
\Big(D\vec{\mathcal{B}}\Big)= \bf{1}, \label{e23}
\end{eqnarray}
where $D$ matrix is given by Eq. (\ref{e13}),
$\vec{\mathcal{B}}=T_{{}^{6}He}{1\choose 0}={\mathcal{B}_{0}\choose
\mathcal{B}_{1}}$ is the bound state vector, and $K$ is given by
$K(E,q,q')=R(E,q,q',q,q')+ H(q,q',\Lambda)$. We must define the
inverse propagators matrix $I=\mathrm{diag}(I_{0},I_{1})$ with
\begin{eqnarray}
I_{0,1}= \frac{2\pi^{2}}{q^{2}}\;\delta(q-q')\;
\overline{D}_{0,1}(E,q)^{-1}. \label{e24}
\end{eqnarray}

We have defined here the short notation \cite{h7}
 \begin{eqnarray}
A\otimes B \equiv \frac{1}{2 \pi^{2}} \int_{0}^{\Lambda} q^{2} dq\;
A(\cdots, q) B(q,\cdots). \label{e25}
\end{eqnarray}
Therefore for each cutoff $\Lambda$, we have fixed the
$H_{0}(\Lambda)=\frac{h^{2}}{\Omega}$ parameter such that Eq.
(\ref{e21}) is satisfied at experimental value of $E=-B_{2n}=-0.97
\;\textrm{MeV}$.


\section{The contribution of diagrams $(\textrm{a})$, $(\textrm{b})$, $(\textrm{c})$ and $(\textrm{d})$ to charge form factor} \label{Appendix B}
In this appendix, we introduce explicitly the relations of four
different diagrams (a), (b), (c) and (d) that contribute to the charge
form factor as shown in Fig. \ref{fig5}.
\subsection{Contribution $\mathcal{F}_{E}^{(a)}$}\label{subsection1}
After performing the energy integral analytically, using Eqs.
(\ref{e4}), (\ref{e35})-(\ref{a1}) and Feynman rules, the
contribution of diagram (a) in Fig. \ref{fig5} is given by
\begin{eqnarray}
\mathcal{F}^{(a)}(Q^{2})&=&\int_{0}^{\Lambda}
\frac{p^{2}}{2\pi^{2}}\;dp\;\int_{0}^{\Lambda}
\frac{k^{2}}{2\pi^{2}}\;dk\;\vec{\mathcal{G}}^{irr}(p)^{T}
\nonumber\\
&&\times D(p)\;\Upsilon^{(a)}(Q,p,k)
D(k)\;\vec{\mathcal{G}}^{irr}(k), \label{e42}
\end{eqnarray}
where the components of $2\times2$ matrix $\Upsilon^{(a)}(Q,p,k)$
are given by
\begin{eqnarray}
&&\Upsilon^{(a)}_{ij}(Q,p,k)=\frac{g_{1}^{2}}{4}\;\frac{(2\mu)^{2}}{8\pi}\int_{-1}^{1}dx\int_{-1}^{1}dy\int_{0}^{2\pi}d\phi\nonumber\\
&&\times\bigg \{p^{2}+k^{2}+r'^{2}-2(ky+\frac{m_{n}}{M_{1}}px)r'
\nonumber\\
&&\quad+\frac{2m_{n}}{M_{1}}k p(\sqrt{1-x^{2}}\sqrt{1-y^{2}}\cos\phi+xy)+2\mu B_{2n}\bigg\}^{-1}\nonumber\\
&&\times\bigg\{p^{2}+k^{2}+r'^{2}+2(px+\frac{m_{n}}{M_{1}}ky)r'
\nonumber\\
&&\quad+\frac{2m_{n}}{M_{1}}k p(\sqrt{1-x^{2}}\sqrt{1-y^{2}}\cos\phi+xy)+2\mu B_{2n}\bigg\}^{-1}\nonumber\\
&&\times U(p, k, Q, x, y, \phi)\;\delta_{i1}\;\delta_{j1},
\label{e43}
\end{eqnarray}
where $\vec{r'}=\frac{m_{n}}{M_{tot}}\vec{Q}$,
$\mu=\frac{m_{n}m_{\alpha}}{m_{n}+m_{\alpha}}$, the polar angles\\
$x=\cos(\angle(\vec{Q},\vec{p}))$,
$y=\cos(\angle(\vec{Q},\vec{k}))$,\\
$\cos(\angle(\vec{p},\vec{k}))=\sqrt{(1-x^{2})}\sqrt{(1-y^{2})}\cos\phi+xy$
and
\begin{eqnarray}
&&U(p, k, Q, x, y, \phi)=-4 r'^{2}x y+\frac{4}{3}p y
r'\Big(1+\frac{4 m_{n}}{M_{1}}\Big)
\nonumber\\&&-\frac{4}{3}k x r'\Big(1+\frac{4 m_{n}}{M_{1}}\Big)\nonumber\\
&&\quad-\frac{20}{3}p x r'(\sqrt{1-x^{2}}\sqrt{1-y^{2}}\cos\phi+xy)\nonumber\\
&&\quad+\frac{20}{3}k y r'(\sqrt{1-x^{2}}\sqrt{1-y^{2}}\cos\phi+xy)\nonumber\\
&&\quad+\frac{16}{3}\frac{m_{n}}{M_{1}}\;p^{2}(\sqrt{1-x^{2}}\sqrt{1-y^{2}}\cos\phi+xy)
\nonumber\\ &&\quad+\frac{16}{3}\frac{m_{n}}{M_{1}}\;k^{2}(\sqrt{1-x^{2}}\sqrt{1-y^{2}}\cos\phi+xy)\nonumber\\
&&\quad-\frac{4}{3}\;r'^{2}(\sqrt{1-x^{2}}\sqrt{1-y^{2}}\cos\phi+xy)\nonumber\\
&&\quad+\frac{20}{3}\;k
p(\sqrt{1-x^{2}}\sqrt{1-y^{2}}\cos\phi+xy)^{2}
\nonumber\\&&\quad-\frac{4}{3} p k\Big(1-\frac{4
m_{n}^{2}}{M_{1}^{2}}\Big). \label{e44}
\end{eqnarray}

\subsection{Contribution $\mathcal{F}_{E}^{(b)}$}\label{subsection2}
For the contribution of the diagram (b) in Fig. \ref{fig5},
calculating the energy integral analytically, applying Eqs.
(\ref{e4}), (\ref{e35})-(\ref{a1}) and Feynman rules lead to the
following relations
\begin{eqnarray}
\mathcal{F}^{(b)}(Q^{2})&=&\;\int_{0}^{\Lambda}
\Big(-\frac{p^{2}}{2\pi^{2}}\Big)\;dp\;\int_{0}^{\Lambda}
\Big(-\frac{k^{2}}{2\pi^{2}}\Big)\;dk
\nonumber\\&\times&\vec{\mathcal{G}}^{irr}(p)^{T}\;D(p)
\;\Upsilon^{(b)}(Q,p,k)\;D(k)\;\vec{\mathcal{G}}^{irr}(k),\nonumber\\
\label{e45}
\end{eqnarray}
where
\begin{eqnarray}
\Upsilon^{(b)}_{ij}(Q,p,k)=\!\frac{m_{\alpha}}{2}\!\int_{0}^{\Lambda}
\!\Big(\frac{q^{2}}{2\pi^{2}}\!\Big)\!\frac{d
q}{q}\!\int_{-1}^{1}\!\frac{d x'}{x'}\chi^{(b)}_{ij}(r', q,
x',p,k),\nonumber\\ \label{e46}
\end{eqnarray}
so that
\begin{eqnarray}
&&\chi^{(b)}_{ij}(r', q, x', p, k)=\;\frac{1}{Q}\;\Big\{R_{i0}\Big(-B_{2n}, p, d(q, r', x')\Big)\nonumber\\
&&\quad\times\overline{D}_{0}\Big(-B_{2n}\;-\;\frac{m_{n}}{4m_{\alpha}M_{tot}}\;Q^2\;-\;\frac{q Q x'}{2 m_{\alpha}}, q\Big)\nonumber\\
&&\quad\times R_{0j}\Big(-B_{2n}\;-\;\frac{q Q x'}{m_{\alpha}}, d(q, r', -x'), k\Big)\nonumber\\
&&-R_{i0}\Big(-B_{2n}\;+\;\frac{q Q x'}{m_{\alpha}}, p, d(q, r', x')\Big)\nonumber\\
&&\quad\times\overline{D}_{0}\Big(-B_{2n}\;-\;\frac{m_{n}}{4m_{\alpha}M_{tot}}\;Q^2\;+\;\frac{q Q x'}{2 m_{\alpha}}, q\Big)\nonumber\\
&&\quad\times R_{0j}\Big(-B_{2n}, d(q, r',-x'),
k\Big)\Big\}\;\delta_{i1}\;\delta_{j1}, \label{e47}
\end{eqnarray}
with $ d(q, r', \pm x')=\sqrt{q^{2}+r'^{2}\pm 2qr'x'}$.

\subsection{Contribution $\mathcal{F}_{E}^{(c)}$}\label{subsection3}
The leading contribution to charge form factor in the
${}^{6}\textrm{He}$ halo nucleus comes from diagram (c) in Fig.
\ref{fig5} by coupling the photon to $\alpha$ core inside a
$n\alpha$ bubble.
\begin{figure}
\centering
\includegraphics[width=8cm]{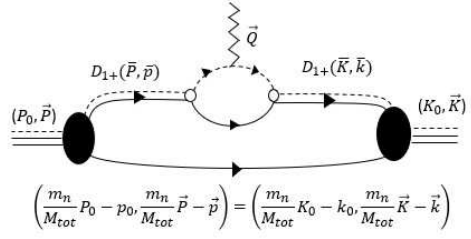}
 \caption{\small{Diagrammatic representation of diagram (c).}}\label{fig6}
\end{figure}
For calculating the contribution of the form factor presented in
Fig. \ref{fig6}, we start with the four-momentum integration as
\begin{eqnarray}
&&\mathcal{F}^{(c)}(Q^{2})=(-ie\mathcal{Z})^{-1}\;\int_{k<\Lambda}\frac{d^{4}k}{(2\pi)^{4}}
\;\int_{p<\Lambda} \frac{d^{4}p}{(2\pi)^{4}}\;(2\pi)^{4}
\nonumber\\&&\times\delta(k_{0}-p_{0})\delta^{(3)}(\vec{k}-\vec{p}-\frac{m_{n}}{M_{tot}}\vec{Q})
\nonumber\\
&&\times(-ie\mathcal{Z})\;
\Big\{i\;\mathcal{G}_{1}^{irr}(P_{0},\vec{P},p_{0},\vec{P})^{T}
\nonumber\\&&\qquad\qquad\times\frac{i}{\frac{m_{n}}{M_{tot}}P_{0}-p_{0}-\frac{(\frac{m_{n}}{M_{tot}}\vec{P}-\vec{p})^{2}}{2m_{n}}+i\varepsilon}
\nonumber\\
&&\qquad\qquad\times\Big[-i\Sigma\Big(\frac{M_{1}}{M_{tot}}\bar{P}+\bar{p},\frac{M_{1}}{M_{tot}}\bar{K}+\bar{k}\Big)\Big]
\nonumber\\&&\qquad\qquad\times i D_{1}\Big(\frac{M_{1}}{M_{tot}}K_{0}+k_{0},\frac{M_{1}}{M_{tot}}\vec{K}+\vec{k}\Big)\nonumber\\
&&\qquad\qquad\times i
D_{1}\Big(\frac{M_{1}}{M_{tot}}P_{0}+p_{0},\frac{M_{1}}{M_{tot}}\vec{P}+\vec{p}\Big)
\nonumber\\&&\qquad\qquad\times
i\;\mathcal{G}_{1}^{irr}(K_{0},\vec{K},k_{0},\vec{k})\Big\},\label{a3}
\end{eqnarray}
where
$\Sigma\Big(\frac{M_{1}}{M_{tot}}\bar{P}+\bar{p},\frac{M_{1}}{M_{tot}}\bar{K}+\bar{k}\Big)$
is the $n\alpha$ bubble contribution in Fig. \ref{fig6}. One of the
two four-momentum integrations in Eq. (\ref{a3}) is absorbed by a
delta-function, so we obtain
\begin{eqnarray}
\mathcal{F}^{(c)}(Q^{2})&=&
\int_{p<\Lambda}\frac{d^{4}p}{(2\pi)^{4}}\;
\Big\{\mathcal{G}_{1}^{irr}(P_{0},\vec{P},p_{0},\vec{P})^{T}
\nonumber\\
&\times&\frac{i}{\frac{m_{n}}{M_{tot}}P_{0}-p_{0}-\frac{(\frac{m_{n}}{M_{tot}}\vec{P}-\vec{p})^{2}}{2m_{n}}+i\varepsilon}
\nonumber\\
&\times&\Big[-i\Sigma\Big(\frac{M_{1}}{M_{tot}}\bar{P}+\bar{p},\frac{M_{1}}{M_{tot}}\bar{K}+\bar{k}\Big)\Big]
\nonumber\\
&\times&
D_{1}\Big(\frac{M_{1}}{M_{tot}}K_{0}+p_{0},\frac{M_{1}}{M_{tot}}\vec{K}+\vec{p}+\frac{m_{n}}{M_{tot}}\vec{Q}\Big)
\nonumber\\
&\times&
D_{1}\Big(\frac{M_{1}}{M_{tot}}P_{0}+p_{0},\frac{M_{1}}{M_{tot}}\vec{P}+\vec{p}\Big)
\nonumber\\
&\times&\mathcal{G}_{1}^{irr}(K_{0},\vec{K},p_{0},\vec{p}+\frac{m_{n}}{M_{tot}}\vec{Q})\Big\}.
\label{a4}
\end{eqnarray}
Using the rescaled four-momentum
$\bar{s}=\frac{m_{n}}{2M_{tot}}\bar{Q}$ and the shifted loop
momentum according to $\bar{p}\mapsto\bar{q}-\bar{s}$, we have
\begin{eqnarray}
\mathcal{F}^{(c)}(Q^{2})&=&\int_{q<\Lambda}\frac{d^{3}\vec{q}}{(2\pi)^{3}}\int_{-\infty}^{+\infty}\frac{d
q_{0}}{2\pi}\;
\Big\{\mathcal{G}_{1}^{irr}(P_{0},\vec{P},\bar{q}-\bar{s})^{T}
\nonumber\\
&\times&\frac{i}{\frac{m_{n}}{M_{tot}}P_{0}-q_{0}-\frac{(\frac{m_{n}}{M_{tot}}\vec{P}-\vec{q}+\vec{s})^{2}}{2m_{n}}
+i\varepsilon}
\nonumber\\
&\times&\Big[-i\Sigma\Big(\frac{M_{1}}{M_{tot}}\bar{P}+\bar{q}-\bar{s},\frac{M_{1}}{M_{tot}}\bar{K}+\bar{q}+\bar{s}\Big)\Big]
\nonumber\\
&\times& D_{1}
\Big(\frac{M_{1}}{M_{tot}}K_{0}+q_{0},\frac{M_{1}}{M_{tot}}\vec{K}+\vec{q}+\vec{s}\Big)\nonumber\\
&\times&
D_{1}\Big(\frac{M_{1}}{M_{tot}}P_{0}+q_{0},\frac{M_{1}}{M_{tot}}\vec{P}+\vec{q}-\vec{s}\Big)
\nonumber\\
&\times&\mathcal{G}_{1}^{irr}(K_{0},\vec{K},\bar{q}+\bar{s})\Big\}.
\label{a5}
\end{eqnarray}
After performing the $q_{0}$ integration according to the pole
$q_{0}=\frac{m_{n}}{M_{tot}}P_{0}-\frac{(\frac{m_{n}}{M_{tot}}\vec{P}-\vec{q}+\vec{s})^{2}}{2m_{n}}+i\varepsilon$,
we obtain
\begin{eqnarray}
\mathcal{F}^{(c)}(Q^{2})&=&\int_{q<\Lambda}\frac{d^{3}\vec{q}}{(2\pi)^{3}}\;
\Big\{\mathcal{G}_{1}^{irr}(P_{0},\vec{P},\bar{q}-\bar{s})^{T}\;
\nonumber\\&\times&\Big[-i\Sigma\Big(\frac{M_{1}}{M_{tot}}\bar{P}+\bar{q}-\bar{s},\frac{M_{1}}{M_{tot}}\bar{K}+\bar{q}+\bar{s}\Big)\Big]
\nonumber\\
&\times& D_{1}\Big(\frac{M_{1}}{M_{tot}}K_{0}+q_{0},\frac{M_{1}}{M_{tot}}\vec{K}+\vec{q}+\vec{s}\Big)\nonumber\\
&\times&
D_{1}\Big(\frac{M_{1}}{M_{tot}}P_{0}+q_{0},\frac{M_{1}}{M_{tot}}\vec{P}+\vec{q}-\vec{s}\Big)
\nonumber\\
&\times&\mathcal{G}_{1}^{irr}(K_{0},\vec{K},\bar{q}+\bar{s})\Big\}.\label{a6}
\end{eqnarray}
Considering Eq. (\ref{a1}) and by substituting the pole $q_{0}$, we
have
\begin{eqnarray}
\mathcal{G}_{1}^{irr}(P_{0},\vec{P},\bar{q}-\bar{s})^{T}&=&\frac{1}{\sqrt{3}}\;\frac{m_{n}\;g_{1}^{2}}{\Lambda^{2}}\;\left|\beta\;H_{0}(\Lambda)\right|\;d(q, s, -x')\nonumber\\
&-&\sum_{i=0}^{1}\int_{0}^{\Lambda}dp\frac{p^{2}}{2\pi^{2}}\;\mathcal{G}_{i}^{irr}(E,p)\;\overline{D}_{i}(E,p)
\nonumber\\
&&\quad\times R_{i1}\Big(E,p,d(q, s, -x')\Big),\label{a7}
\end{eqnarray}
\begin{eqnarray}
\mathcal{G}_{1}^{irr}(K_{0},\vec{K},\bar{q}+\bar{s})&=&\frac{1}{\sqrt{3}}\;\frac{m_{n}\;g_{1}^{2}}{\Lambda^{2}}\;\left|\beta\;H_{0}(\Lambda)\right|
\;d(q, s, x')\nonumber\\
&-&\sum_{j=0}^{1}\int_{0}^{\Lambda}dk\frac{k^{2}}{2\pi^{2}}\;R_{1j}\Big(E,d(q,
s,
x'),k\Big)\nonumber\\
&&\quad\times\overline{D}_{j}(E,k)\;\mathcal{G}_{j}^{irr}(E,k),\label{a8}
\end{eqnarray}
with $d(q, s, \pm x')=\sqrt{q^{2}+s^{2}\pm 2qsx'}$ and the polar
angle $x'=\cos(\angle(\vec{Q},\vec{q}))$. Using Eq. (\ref{e9}), Eq.
(\ref{e13}) and inserting the pole $q_{0}$, we can redefine the two
propagators in Eq. (\ref{a5}) as
\begin{eqnarray}
&&D_{1}\Big(\frac{M_{1}}{M_{tot}}K_{0}+q_{0},\frac{M_{1}}{M_{tot}}\vec{K}+\vec{q}+\vec{s}\Big)
\nonumber\\&&\qquad\quad\qquad
=\overline{D}_{1}\Big(E-\frac{m_{n}}{8M_{1}M_{tot}}Q^{2}-\frac{q Q
x'}{2M_{1}},q\Big),\label{a9}\nonumber\\
\end{eqnarray}
\begin{eqnarray}
&&D_{1}\Big(\frac{M_{1}}{M_{tot}}P_{0}+q_{0},\frac{M_{1}}{M_{tot}}\vec{P}+\vec{q}-\vec{s}\Big)
\nonumber\\&&\qquad\quad\qquad=\overline{D}_{1}\Big(E-\frac{m_{n}}{8M_{1}M_{tot}}Q^{2}+\frac{q
Q x'}{2M_{1}},q\Big).\label{a10}\nonumber\\
\end{eqnarray}

\subsubsection{Bubble diagram}\label{Bubble diagram}
\begin{figure}
\centering
\includegraphics[width=6cm]{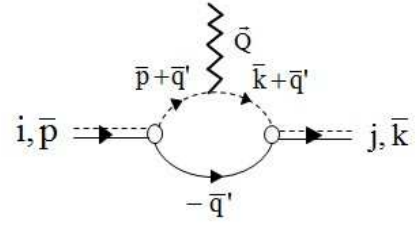}
 \caption{\small{Diagrammatic representation of the bubble contribution $-i\Sigma(\bar{p},\bar{k})$.}}\label{fig7}
\end{figure}
We now calculate the term
$-i\Sigma\Big(\frac{M_{1}}{M_{tot}}\bar{P}+\bar{q}-\bar{s},\frac{M_{1}}{M_{tot}}\bar{K}+\bar{q}+\bar{s}\Big)$
for the bubble diagram depicted in Fig. \ref{fig7}. For general
incoming (outgoing) four-momenta $\bar{p}\;(\bar{k})$ and according
to Eq. (\ref{e4}), we obtain
\begin{eqnarray}
-i\Sigma(\bar{p},\bar{k})&=&-\int\frac{d^{3}\vec{q'}}{(2\pi)^{3}}\Big(i\frac{g_{1}}{2}\Big)^{2}
\nonumber\\&\times&\Big[\frac{1}{\frac{q'^{2}}{2m_{n}}+\frac{(\vec{k}+\vec{q'})^{2}}{2m_{\alpha}}-k_{0}}\times
\frac{1}{\frac{q'^{2}}{2m_{n}}+\frac{(\vec{p}+\vec{q'})^{2}}{2m_{\alpha}}-p_{0}}\nonumber\\&\times&
(\vec{k}(1-r)+2\vec{q'})_{j}(\vec{p}(1-r)+2\vec{q'})_{i}\;\textrm{Tr}(S^{\dag}_{j}S_{i})\Big].\nonumber\\\label{a11}
\end{eqnarray}
According to Eq. (\ref{e6}), we derive
$\textrm{Tr}(S^{\dag}_{j}S_{i})=\frac{4}{3}\delta_{ij}$.
Furthermore, using the relation
$\frac{1}{a_{1}a_{2}}=\int_{0}^{1}\frac{dx}{[a_{1}x+a_{2}(1-x)]^{2}}$
and defining the rescaled loop momentum
$\vec{b}:=\frac{\vec{q'}+\frac{\mu}{m_{\alpha}}[x\vec{p}+(1-x)\vec{k}]}{\frac{\mu}{m_{\alpha}}Q}$
and replacing our kinematics
$\bar{p}\mapsto\frac{M_{1}}{M_{tot}}\bar{P}+\bar{q}-\bar{s}$ and
$\bar{k}\mapsto\frac{M_{1}}{M_{tot}}\bar{K}+\bar{q}+\bar{s}$ in Eq.
(\ref{a11}), we can obtain the following relation for the bubble
contribution, $\Sigma$, as
\begin{eqnarray}
&-&i\Sigma\Big(\frac{M_{1}}{M_{tot}}\bar{P}+\bar{q}-\bar{s},\frac{M_{1}}{M_{tot}}\bar{K}+\bar{q}+\bar{s}\Big)
\nonumber\\
&&=\frac{4}{3}\frac{\mu\;m_{\alpha}}{Q}g_{1}^{2}\int_{0}^{1}dx\int\frac{d^{3}\vec{b}}{(2\pi)^{3}}
\Bigg\{\frac{1}{(b^{2}-A_{\left|\overline{q}-\overline{s}\right|,\left|\overline{q}+\overline{s}\right|}(x))^{2}}
\nonumber\\
&&\times\Bigg[\Big(\frac{M_{1}}{M_{tot}}\vec{K}+\vec{q}+\vec{s}\Big)(1-r)
\nonumber\\&&\qquad+
\frac{2\mu}{m_{\alpha}}\Big(Q\vec{b}-\Big[x\Big(\frac{M_{1}}{M_{tot}}\vec{P}+\vec{q}-\vec{s}\Big)
\nonumber\\&&\qquad+(1-x)\Big(\frac{M_{1}}{M_{tot}}\vec{K}+\vec{q}+\vec{s}\Big)\Big]\Big)\Bigg]_{i}
\nonumber\\
&&\times\Bigg[\Big(\frac{M_{1}}{M_{tot}}\vec{P}+\vec{q}-\vec{s}\Big)(1-r)
\nonumber\\&&\qquad+
\frac{2\mu}{m_{\alpha}}\Big(Q\vec{b}-\Big[x\Big(\frac{M_{1}}{M_{tot}}\vec{P}+\vec{q}-\vec{s}\Big)
\nonumber\\&&\qquad+(1-x)\Big(\frac{M_{1}}{M_{tot}}\vec{K}+\vec{q}+\vec{s}\Big)\Big]\Big)\Bigg]_{i}\Bigg\},
 \label{a12}
\end{eqnarray}
where
\begin{eqnarray}
A_{\left|\overline{q}-\overline{s}\right|,\left|\overline{q}+\overline{s}\right|}(x)=x^{2}-
x\Big(1+C_{\overline{P},\overline{q}-\overline{s}}-
C_{\overline{K},\overline{q}+\overline{s}}\Big)-C_{\overline{K},\overline{q}+\overline{s}},\nonumber\\
\label{a13}
\end{eqnarray}
with
\begin{eqnarray}
C_{\overline{P},\overline{q}-\overline{s}}=2\mu\Big[B_{2n}+\frac{(\vec{q}-\vec{s})^{2}}{2\tilde{\mu}}\Big]\Big(\frac{M_{1}}{2M_{tot}}\Big)^{2}\frac{1}{s^{2}},\nonumber\\
C_{\overline{K},\overline{q}+\overline{s}}=2\mu\Big[B_{2n}+\frac{(\vec{q}+\vec{s})^{2}}{2\tilde{\mu}}\Big]\Big(\frac{M_{1}}{2M_{tot}}\Big)^{2}\frac{1}{s^{2}},
\label{a14}
\end{eqnarray}
and $\tilde{\mu}=\frac{m_{n} M_{1}}{M_{tot}}$. Finally, after some
derivations, the relation of the bubble contribution in
Eq.~(\ref{a12}) converts to
\begin{eqnarray}
&-&i\Sigma\Big(\frac{M_{1}}{M_{tot}}\bar{P}+\bar{q}-\bar{s},\frac{M_{1}}{M_{tot}}\bar{K}+\bar{q}+\bar{s}\Big)
\nonumber\\&=&\frac{4}{3}\frac{\mu
m_{\alpha}}{Q}g_{1}^{2}\int_{0}^{1}dx\Bigg\{4\frac{\mu^{2}}{m_{\alpha}^{2}}Q^{2}
\int
\frac{d^{3}\vec{b}}{(2\pi)^{3}}\frac{b^{2}}{[b^{2}-A_{\left|\overline{q}-\overline{s}\right|,\left|\overline{q}+\overline{s}\right|}(x)]^{2}}
\nonumber\\
&&+16Qs\Big(\frac{\mu^{2}}{m_{\alpha}^{2}}\frac{M_{1}}{M_{tot}}\Big)\!(x^{2}\!-\!x)\!\int\!\frac{d^{3}\vec{b}}{(2\pi)^{3}}
\frac{1}{[b^{2}-A_{\left|\overline{q}-\overline{s}\right|,\left|\overline{q}+\overline{s}\right|}(x)]^{2}}
\nonumber\\&&+
16\frac{\mu^{2}}{m_{\alpha}^{2}}s^{2}(\!x^{2}\!-x\!)\!\int\!\frac{d^{3}\vec{b}}{(2\pi)^{3}}
\frac{1}{[b^{2}-A_{\left|\overline{q}-\overline{s}\right|,\left|\overline{q}+\overline{s}\right|}(x)]^{2}}
\nonumber\\
&&+4\Big(\frac{\mu^{2}}{m_{\alpha}^{2}}\frac{M_{1}^{2}}{M_{tot}^{2}}\Big)Q^{2}(\!x^{2}\!-\!x\!)\!\int\!\frac{d^{3}\vec{b}}{(2\pi)^{3}}
\frac{1}{[b^{2}-A_{\left|\overline{q}-\overline{s}\right|,\left|\overline{q}+\overline{s}\right|}(x)]^{2}}\Bigg\}
\nonumber\\
&=&\frac{2i}{3\pi}\frac{\mu^{3}}{m_{\alpha}}g_{1}^{2}Q[-3I_{4}+I_{3}-I_{2}],
\label{a016}
\end{eqnarray}
where the functions $I_2$, $I_3$ and $I_4$ in the last line are
given using the relations
\begin{eqnarray}
\int\frac{d^{3}\vec{b}}{(2\pi)^{3}}
\frac{1}{[b^{2}-A_{\left|\overline{q}-\overline{s}\right|,\left|\overline{q}+\overline{s}\right|}(x)]^{2}}=
\frac{i}{8\pi}\frac{1}{\sqrt{A_{\left|\overline{q}-\overline{s}\right|,\left|\overline{q}+\overline{s}\right|}(x)}},\nonumber\\
\label{a0017}
\end{eqnarray}
and
\begin{eqnarray}
\int\frac{d^{3}\vec{b}}{(2\pi)^{3}}
\frac{b^{2}}{[b^{2}-A_{\left|\overline{q}-\overline{s}\right|,\left|\overline{q}+\overline{s}\right|}(x)]^{2}}=
\frac{3i}{8\pi}\sqrt{A_{\left|\overline{q}-\overline{s}\right|,\left|\overline{q}+\overline{s}\right|}(x)},\nonumber\\
\label{a0018}
\end{eqnarray}
as
\begin{eqnarray}
I_{2}&=&\int_{0}^{1}dx\frac{x}{\sqrt{A_{\left|\overline{q}-\overline{s}\right|,\left|\overline{q}+\overline{s}\right|}(x)}}
\nonumber\\
&=&i\Big(\sqrt{C_{\overline{P},\overline{q}-\overline{s}}}-\sqrt{C_{\overline{K},\overline{q}+\overline{s}}}\Big)
\!+\!\frac{(1+C_{\overline{P},\overline{q}-\overline{s}}-C_{\overline{K},\overline{q}+\overline{s}})}{2}I_{1},
\nonumber\\\label{a019}
\\
I_{3}&=&\int_{0}^{1}dx\frac{x^{2}}{\sqrt{A_{\left|\overline{q}-\overline{s}\right|,\left|\overline{q}+\overline{s}\right|}(x)}}
\nonumber\\
&=&\frac{1}{4}\Big[(1+C_{\overline{K},\overline{q}+\overline{s}}-C_{\overline{P},\overline{q}-\overline{s}})
i\sqrt{C_{\overline{P},\overline{q}-\overline{s}}}
\nonumber\\&&+(1+C_{\overline{P},\overline{q}-\overline{s}}-C_{\overline{K},\overline{q}+\overline{s}})
i\sqrt{C_{\overline{K},\overline{q}+\overline{s}}}\;\Big]
\nonumber\\
&&+\frac{1}{2}\Big(C_{\overline{K},\overline{q}+\overline{s}}+\frac{(1+C_{\overline{P},\overline{q}-\overline{s}}
-C_{\overline{K},\overline{q}+\overline{s}})^{2}}{4}\Big)I_{1}
\nonumber\\
&&+(1+C_{\overline{P},\overline{q}-\overline{s}}-C_{\overline{K},\overline{q}+\overline{s}})i\Big(\sqrt{C_{\overline{P},\overline{q}-\overline{s}}}-\sqrt{C_{\overline{K},\overline{q}+\overline{s}}}\;\Big)
\nonumber\\&&+\frac{(1+C_{\overline{P},\overline{q}-\overline{s}}-C_{\overline{K},\overline{q}+\overline{s}})^{2}}{4}I_{1},
\end{eqnarray}
\begin{eqnarray}
I_{4}&=&\int_{0}^{1} dx \sqrt{A_{\left|\overline{q}-\overline{s}\right|,\left|\overline{q}+\overline{s}\right|}(x)}\nonumber\\
&=&\frac{1}{4}\Big[(1+C_{\overline{K},\overline{q}+\overline{s}}-C_{\overline{P},\overline{q}-\overline{s}})i\sqrt{C_{\overline{P},\overline{q}-\overline{s}}}
\nonumber\\&&+(1+C_{\overline{P},\overline{q}-\overline{s}}-C_{\overline{K},\overline{q}+\overline{s}})i\sqrt{C_{\overline{K},\overline{q}+\overline{s}}}\;\Big]\nonumber\\
&&-\frac{1}{2}\Big(C_{\overline{K},\overline{q}+\overline{s}}+\frac{(1+C_{\overline{P},\overline{q}-\overline{s}}-C_{\overline{K},\overline{q}+\overline{s}})^{2}}{4}\Big)I_{1}.
\label{a020}\nonumber\\
\end{eqnarray}
The $I_1$ function in Eqs.~(\ref{a019})-(\ref{a020}) is defined by
the expression
\begin{eqnarray}
I_{1}&=&\int_{0}^{1}\frac{dx}{\sqrt{A_{\left|\overline{q}-\overline{s}\right|,\left|\overline{q}+\overline{s}\right|}(x)}}\nonumber\\
&=&-i\Bigg[\arctan\Bigg(\frac{\frac{M_{tot}}{M_{1}}s+\frac{m_{\alpha}}{M_{1}}qx'}{\sqrt{2\mu\Big(B_{2n}+\frac{(\vec{q}-\vec{s})^{2}}{2\tilde{\mu}}\Big)}}\Bigg)
\nonumber\\&&\qquad+\arctan\Bigg(\frac{\frac{M_{tot}}{M_{1}}s-\frac{m_{\alpha}}{M_{1}}qx'}{\sqrt{2\mu\Big(B_{2n}+\frac{(\vec{q}+\vec{s})^{2}}{2\tilde{\mu}}\Big)}}\Bigg)\Bigg].
\label{a017}
\end{eqnarray}

\subsubsection{Final representation of ${F}^{(c)}(Q^{2})$}\label{Final representation}
In the final step, by inserting Eqs. (\ref{a7})-(\ref{a10}) into Eq.
(\ref{a6}), we find
\begin{eqnarray}
&&\mathcal{F}^{(c)}(Q^{2})=\;\int_{0}^{\Lambda} \Big(-\frac{p^{2}}{2\pi^{2}}\Big)\;dp\;\int_{0}^{\Lambda}
\Big(-\frac{k^{2}}{2\pi^{2}}\Big)\;dk\;\vec{\mathcal{G}}^{irr}(p)^{T}
\nonumber\\&&\times D(p)\;\Upsilon^{(c)}(Q,p,k)\;D(k)\;\vec{\mathcal{G}}^{irr}(k)\nonumber\\
&&+2\int_{0}^{\Lambda}
\Big(-\frac{p^{2}}{2\pi^{2}}\Big)dp\vec{\mathcal{G}}^{irr}(p)^{T}D(p)\vec{\Upsilon}^{(c)}(Q,p)+\Upsilon^{(c)}_{0}(Q),\nonumber\\
\label{ea13}
\end{eqnarray}
where
\begin{eqnarray}
\Upsilon^{(c)}_{ij}(Q,p,k)&=&\frac{1}{(2\pi)^{3}}\;\int_{0}^{\Lambda}q^{2}\;dq\;\int_{-1}^{1}dx'\;\int_{0}^{2\pi}d\phi
\nonumber\\&&\times\chi^{(c)}_{ij}(\frac{m_{n}}{2M_{tot}}Q, p, k, q, x', \phi ),\nonumber\\
\Upsilon^{(c)}_{i}(Q,p)&=&\frac{1}{(2\pi)^{3}}\;\int_{0}^{\Lambda}q^{2}\;dq\;\int_{-1}^{1}dx'\;\int_{0}^{2\pi}d\phi
\nonumber\\&&\times\chi^{(c)}_{i}(\frac{m_{n}}{2M_{tot}}Q, p, q, x', \phi ),\nonumber\\
\Upsilon^{(c)}_{0}(Q)&=&\frac{1}{(2\pi)^{3}}\int_{0}^{\Lambda}q^{2}dq\int_{-1}^{1}dx'\int_{0}^{2\pi}d\phi
\nonumber\\&&\times\chi^{(c)}_{0}(\frac{m_{n}}{2M_{tot}}Q, q, x',
\phi ), \label{ea14}
\end{eqnarray}
with the following relations
\begin{eqnarray}
&&\chi^{(c)}_{ij}(\frac{m_{n}}{2M_{tot}}Q, q, x, p, k)=R_{i1}\Big(-B_{2n}, p, d(q, s, -x')\Big)\nonumber\\
&&\qquad\times\overline{D}_{1}\Big(-B_{2n}\;-\;\frac{m_{n}}{8M_{1}M_{tot}}\;Q^2\;-\;\frac{q Q x'}{2 M_{1}}, q\Big)\nonumber\\
&&\qquad\times\Big[-i\;\Sigma\Big(\frac{M_{1}}{M_{tot}}\bar{P}+\bar{q}-\bar{s},\frac{M_{1}}{M_{tot}}\bar{K}+\bar{q}+\bar{s}\Big)\Big]\nonumber\\
&&\qquad\times\overline{D}_{1}\Big(-B_{2n}\;-\;\frac{m_{n}}{8M_{1}M_{tot}}\;Q^2\;+\;\frac{q Q x'}{2 M_{1}}, q\Big)\nonumber\\
&&\qquad\times R_{1j}\Big(-B_{2n}, d(q, s, x'), k\Big), \label{a15}
\end{eqnarray}
\begin{eqnarray}
&&\chi^{(c)}_{i}(\frac{m_{n}}{2M_{tot}}Q, p, q, x', \phi )
=\frac{1}{\sqrt{3}}\;\frac{m_{n}\;g_{1}^{2}}{\Lambda^{2}}\left|\beta
H_{0}(\Lambda)\right|d(q, s, x')
\nonumber\\&&\qquad\times R_{i1}\Big(-B_{2n}, p, d(q, s, -x')\Big)\nonumber\\
&&\qquad\times\overline{D}_{1}\Big(-B_{2n}\;-\;\frac{m_{n}}{8M_{1}M_{tot}}\;Q^2\;-\;\frac{q Q x'}{2 M_{1}}, q\Big)\nonumber\\
&&\qquad\times\Big[-i\;\Sigma\Big(\frac{M_{1}}{M_{tot}}\bar{P}+\bar{q}-\bar{s},\frac{M_{1}}{M_{tot}}\bar{K}+\bar{q}+\bar{s}\Big)\Big]\nonumber\\
&&\qquad\times\overline{D}_{1}\Big(-B_{2n}\;-\;\frac{m_{n}}{8M_{1}M_{tot}}\;Q^2\;+\;\frac{q
Q x'}{2 M_{1}}, q\Big), \label{a015}
\end{eqnarray}
\begin{eqnarray}
&&\chi^{(c)}_{0}(\frac{m_{n}}{2M_{tot}}Q, q, x', \phi )=\frac{1}{\sqrt{3}}\frac{m_{n}g_{1}^{2}}{\Lambda^{2}}\left|\beta H_{0}(\Lambda)\right|d(q, s, -x')\nonumber\\
&&\qquad\times\overline{D}_{1}\Big(-B_{2n}\;-\;\frac{m_{n}}{8M_{1}M_{tot}}\;Q^2\;-\;\frac{q Q x'}{2 M_{1}}, q\Big)\nonumber\\
&&\qquad\times\Big[-i\;\Sigma\Big(\frac{M_{1}}{M_{tot}}\bar{P}+\bar{q}-\bar{s},\frac{M_{1}}{M_{tot}}\bar{K}+\bar{q}+\bar{s}\Big)\Big]\nonumber\\
&&\qquad\times\overline{D}_{1}\Big(-B_{2n}\;-\;\frac{m_{n}}{8M_{1}M_{tot}}\;Q^2\;+\;\frac{q Q x'}{2 M_{1}}, q\Big)\nonumber\\
&&\qquad\times\frac{1}{\sqrt{3}}\;\frac{m_{n}\;g_{1}^{2}}{\Lambda^{2}}\;\left|\beta\;H_{0}(\Lambda)\right|\;d(q,
s, x'). \label{a0015}
\end{eqnarray}
As it mentioned all calculations have been performed in Breit frame,
so we have substituted $E= -B_{2n}$ in Eqs.
(\ref{a15})-(\ref{a0015}) and drop this energy variable in
$\vec{\mathcal{G}}^{irr}$ and matrix $D$ in Eq. (\ref{ea13}).

\subsection{Contribution $\mathcal{F}_{E}^{(d)}$}\label{subsection4}
Diagram (d) in Fig. \ref{fig5} is the same as diagram (c) by
converting the $n\alpha$ bubble to the vertex of photon-$d_{1}$
coupling, therefore the contribution of the diagram (d) is given by
\begin{eqnarray}
\mathcal{F}^{(d)}(Q^{2})&=&\;\int_{0}^{\Lambda} \Big(-\frac{p^{2}}{2\pi^{2}}\Big)\;dp\;\int_{0}^{\Lambda}
\Big(-\frac{k^{2}}{2\pi^{2}}\Big)\;dk
\nonumber\\&&\quad\times\vec{\mathcal{G}}^{irr}(p)^{T}\;D(p)\;\Upsilon^{(d)}(Q,p,k)\;D(k)\;\vec{\mathcal{G}}^{irr}(k)\nonumber\\
&&+2\int_{0}^{\Lambda}
\Big(-\frac{p^{2}}{2\pi^{2}}\Big)\;dp\;\vec{\mathcal{G}}^{irr}(p)^{T}\;D(p)\;\vec{\Upsilon}^{(d)}(Q,p)
\nonumber\\&&+\Upsilon^{(d)}_{0}(Q), \label{e52}
\end{eqnarray}
where
\begin{eqnarray}
\Upsilon^{(d)}_{ij}(Q,p,k)&=&\frac{1}{(2\pi)^{3}}\;\int_{0}^{\Lambda}q^{2}\;dq\;\int_{-1}^{1}dx'\;\int_{0}^{2\pi}d\phi
\nonumber\\&&\quad\times\chi^{(d)}_{ij}(\frac{m_{n}}{2M_{tot}}Q, p, k, q, x', \phi ),\nonumber\\
\vec{\Upsilon}^{(d)}_{i}(Q,p)&=&\frac{1}{(2\pi)^{3}}\;\int_{0}^{\Lambda}q^{2}\;dq\;\int_{-1}^{1}dx'\;\int_{0}^{2\pi}d\phi
\nonumber\\&&\quad\times\chi^{(d)}_{i}(\frac{m_{n}}{2M_{tot}}Q, p, q, x', \phi ),\nonumber\\
\Upsilon^{(d)}_{0}(Q)&=&\frac{1}{(2\pi)^{3}}\;\int_{0}^{\Lambda}q^{2}\;dq\;\int_{-1}^{1}dx'\;\int_{0}^{2\pi}d\phi
\nonumber\\&&\quad\times\chi^{(d)}_{0}(\frac{m_{n}}{2M_{tot}}Q, q,
x', \phi ), \label{e53}
\end{eqnarray}
with the following relations
\begin{eqnarray}
&&\chi^{(d)}_{ij}(\frac{m_{n}}{2M_{tot}}Q, q, x', p, k)=R_{i1}\Big(-B_{2n}, p, d(q, s, -x')\Big)\nonumber\\
&&\qquad\times\overline{D}_{1}\Big(-B_{2n}\;-\;\frac{m_{n}}{8M_{1}M_{tot}}\;Q^2\;-\;\frac{q Q x'}{2 M_{1}}, q\Big)\nonumber\\
&&\qquad\times\overline{D}_{1}\Big(-B_{2n}\;-\;\frac{m_{n}}{8M_{1}M_{tot}}\;Q^2\;+\;\frac{q Q x'}{2 M_{1}}, q\Big)\nonumber\\
&&\qquad\times R_{1j}\Big(-B_{2n}, d(q, s, x'), k\Big),\nonumber\\
&&\chi^{(d)}_{i}(\frac{m_{n}}{2M_{tot}}Q, p, q, x', \phi
)=\frac{1}{\sqrt{3}}\;\frac{m_{n}\;g_{1}^{2}}{\Lambda^{2}}\;\left|\beta\;H_{0}(\Lambda)\right|
\nonumber\\&&\qquad\times d(q, s, x')\;R_{i1}\Big(-B_{2n}, p, d(q, s, -x')\Big)\nonumber\\
&&\qquad\times\overline{D}_{1}\Big(-B_{2n}\;-\;\frac{m_{n}}{8M_{1}M_{tot}}\;Q^2\;-\;\frac{q Q x'}{2 M_{1}}, q\Big)\nonumber\\
&&\qquad\times\overline{D}_{1}\Big(-B_{2n}\;-\;\frac{m_{n}}{8M_{1}M_{tot}}\;Q^2\;+\;\frac{q Q x'}{2 M_{1}}, q\Big),\nonumber\\
&&\chi^{(d)}_{0}(\frac{m_{n}}{2M_{tot}}Q, q, x', \phi
)=\frac{1}{\sqrt{3}}\;\frac{m_{n}\;g_{1}^{2}}{\Lambda^{2}}\;\left|\beta\;H_{0}(\Lambda)\right|
\nonumber\\&&\qquad\times d(q, s, -x')\nonumber\\
&&\qquad\times\overline{D}_{1}\Big(-B_{2n}\;-\;\frac{m_{n}}{8M_{1}M_{tot}}\;Q^2\;-\;\frac{q Q x'}{2 M_{1}}, q\Big)\nonumber\\
&&\qquad\times\overline{D}_{1}\Big(-B_{2n}\;-\;\frac{m_{n}}{8M_{1}M_{tot}}\;Q^2\;+\;\frac{q Q x'}{2 M_{1}}, q\Big)\nonumber\\
&&\qquad\times\frac{1}{\sqrt{3}}\;\frac{m_{n}\;g_{1}^{2}}{\Lambda^{2}}\;\left|\beta\;H_{0}(\Lambda)\right|\;d(q,
s, x'). \label{e54}
\end{eqnarray}

\end{document}